\documentclass[aps, prc, reprint, amsmath, amssymb, superscriptaddress, floatfix, nofootinbib]{revtex4-1} 

\usepackage[pdftex]{graphicx}
\DeclareGraphicsExtensions{.jpg,.png,.pdf}

\usepackage[utf8]{inputenc}
\usepackage{hyperref}
\usepackage{amsmath}
\usepackage{amssymb}
\usepackage{amsfonts}
\usepackage{tabularx}
\usepackage{booktabs}
\usepackage{graphicx}
\usepackage{color}
\usepackage{multirow}
\usepackage[perpage,symbol]{footmisc}
\usepackage{verbatim}
\usepackage{dcolumn}
\usepackage{bm}
\usepackage[inline]{enumitem}
\definecolor{theblue}{RGB}{0,50,230}
\hypersetup{
  colorlinks=true,
  linkcolor=theblue,
  citecolor=theblue,
  urlcolor=theblue
}

\newcommand{\pt}{\ensuremath{p}_{\rm T}}
\newcommand{\raa}{\ensuremath{R}_{\rm AA}}
\newcommand{\vtwo}{\ensuremath{v}_{\rm 2}}
\newcommand{\snn}{\sqrt{s_{\rm NN}}}

\usepackage{lineno}

\begin{document}

\title{Probing the transport properties of Quark-Gluon Plasma via\\ heavy-flavor Boltzmann and Langevin dynamics}

\author{Shuang~Li}
\email{lish@ctgu.edu.cn}
\affiliation{%
College of Science, China Three Gorges University, Yichang 443002, China\\
}%
\affiliation{%
Physics Department and Center for Exploration of Energy and Matter,\\
Indiana University, 2401 N Milo B. Sampson Lane, Bloomington, IN 47408, USA
}%
\affiliation{%
Key Laboratory of Quark and Lepton Physics (MOE), Central China Normal University, Wuhan 430079, China\\
}%
\author{Chaowen~Wang}%
\affiliation{%
College of Science, China Three Gorges University, Yichang 443002, China\\
}%
\author{Renzhuo Wan}%
\affiliation{%
Nano optical material and storage device research center, School of electronic and electrical engineering,\\
Wuhan Textile University, Wuhan 430200, China\\
}%
\author{Jinfeng Liao}%
\email{liaoji@indiana.edu}
\affiliation{%
Physics Department and Center for Exploration of Energy and Matter,\\
Indiana University, 2401 N Milo B. Sampson Lane, Bloomington, IN 47408, USA
}%

\date{\today}

\begin{abstract}
The heavy quark propagation behavior inside the quark-gluon plasma (QGP),
is usually described in terms of the Boltzmann dynamics,
which can be reduced to the Langevin approach by assuming
a small momentum transfer for the scattering processes
between heavy quarks and the QGP constituents.
In this work, the temperature and energy dependence of the
transport coefficients are calculated in the framework of both Boltzmann and Langevin dynamics,
by considering only the elastic scattering processes
to have a better comparison and understanding of these two models.
The extracted transport coefficients are found to be larger in the Boltzmann approach as compared
with the Langevin, in particular in the high temperature and high energy region.
Within each of the two theoretical frameworks, we simulate the charm quark production
and the subsequent evolution processes in relativistic heavy-ion collisions.
We find that the energy loss due to elastic scattering is larger from the Boltzmann dynamics,
resulting in a smaller $\raa$ at high $\pt$ ($\pt\gtrsim10~{\rm GeV}$),
for both the charm quark and heavy-flavor mesons.
Boltzmann model produces systematically larger $\vtwo$, in particular at moderate $\pt$,
meanwhile, it shows a stronger broadening behavior for the relative azimuthal angle between initially
back-to-back generated $c\bar{c}$ pairs in the similar region.
They are mainly induced by the stronger interactions between heavy quarks and the QGP partons in Boltzmann,
which are able to transfer more $\vtwo$ from the medium to the heavy quarks,
as well as to pull more $c\bar{c}$ pairs from high momentum to low momentum. 
By comparing the model calculations with available experimental measurements for D-mesons,
a visible deviation can be observed for both the Boltzmann and Langevin approaches.
The missing inelastic contributions allow reducing the discrepancy with data,
and additionally, the relevant Langevin approach is more favored by the $\raa$ data
while the Boltzmann approach is more favored favor by the $\vtwo$ data.
A simultaneous description of both observables appears challenging for both models.
\end{abstract}


\maketitle

\section{INTRODUCTION}\label{sec:Intro}
In ultrarelativistic collisions of heavy nuclei such as Au or Pb,
an extreme high temperature and energy density environment
can be produced around the collision point,
where allows form a new state of nuclear matter consisting of the deconfined quarks and gluons,
namely the quark-gluon plasma, QGP~\cite{Gyulassy05, Shuryak05}.
To investigate its properties,
the experiments using the Au and Pb as the colliding beams
have been carried at the Relativistic Heavy Ion Collider (RHIC) at BNL
and at the Large Hadron Collider (LHC) at
CERN, respectively, in the past two decades~\cite{Muller12, Jurgen17, Shuryak17}.
The QGP was found to induce the jet quenching,
as well as to exhibit the collective flow behavior among various
probes~\cite{PHENIX04, PHENIX05, STAR05, ALICEV213, ALICERAA13}.
The jet quenching phenomenon is known~\cite{WangPRL92} as
the energy loss of the fast partons traversing the QGP medium,
and it can be investigated by measuring the suppression behavior of the
cross-section of the desired particles produced
in nucleus-nucleus collisions to that in binary-scaled
nucleon-nucleon collisions at the same energy,
which is the so-called nuclear modification factor, $\raa$,
\begin{equation}\label{eq:RAA}
\raa(\pt)=\frac{{\rm d}\sigma_{\rm AA}/{\rm d}\pt}{{\rm d}\sigma_{\rm pp}/{\rm d}\pt}.
\end{equation}
The collective effect can be interpolated as the strong collective expansion of QGP
when its (local) thermal equilibrium state is achieved,
and it can be studied by a Fourier expansion~\cite{Zhang96, Flow97}
of the particle azimuthal distributions with respect to the reaction plane, which
is defined as the plane including impact parameter and beam axis.
Normally, the second coefficient, $\vtwo$,
\begin{equation}\label{eq:v2}
\vtwo(\pt)= \biggr \langle \frac{p_{x}^{2}-p_{y}^{2}}{p_{x}^{2}+p_{y}^{2}} \biggr \rangle,
\end{equation}
is called elliptic flow coefficient,
which allows to describe the anisotropy of the transverse momentum.

Heavy quark (HQ), including charm and bottom,
are of particular interest~\cite{HQQGPBraaten91, HQQGPRapp10, HFSummaryGROUP16, HFSummaryAarts17, HFQM17Greco}
since, due to their large mass,  (1) $m_{\rm Q}\gg\Lambda_{\rm QCD}$,
thus, its initial production can be well described by the
perturbative Quantum ChromoDynamics (pQCD) at the next-to-leading order~\cite{FONLL98, FONLL01, FONLL12}, in particular at high $\pt$;
(2) $m_{\rm Q}\gg T$, resulting in the negligible thermal production of HQ pairs in QGP medium with the temperature reached at RHIC and LHC energies.
In addition, HQ flavor is conserved throughout the interactions with the
surrounding QGP constituents, i.e. gluons and (anti-)light quarks.
Therefore, the initially produced HQ pairs will experience the full evolution of QGP,
and serve as its ideal probes.

During the propagating through the QGP medium,
the HQ dynamics is usually described by the Boltzmann or Langevin model~\cite{RalfSummary08, RalfSummary16}.
For the Boltzmann approach, the evolution of the HQ distribution function
behaves the Boltzmann Transport Equation (BTE),
where the elastic and inelastic scattering processes
between HQs and the quasi-particles of QGP
are quantified by the relevant scattering matrix.
Consequently, it can be given with the help of the perturbative QCD.
Due to large HQ mass and moderate medium temperature,
the typical momentum transfers in interactions, $q\sim gT$,
are assumed small, $gT\ll m_{\rm Q}$~\cite{POWLANGEPJC11},
therefore, the HQ trajectory will be changed significantly only after receiving lots of
soft momentum kicks from the surrounding QGP constituents,
resulting in the Brownian motion.
Based on this assumption, BTE is reduced to the Fokker-Plank Transport Equation (FPTE),
which can be realized stochastically by a Langevin Transport Equation (LTE).
In the framework of LTE, all the interactions are conveniently encoded into three transport coefficients, satisfying the dissipation-fluctuation relation.
Therefore, with LTE, all the problem reduced to the evaluation of three transport coefficients,
which can be extracted from the lattice QCD at zero momentum limit.

Many models were developed from the Boltzmann~\cite{BAMPS10, LBTPRC16, DjordjevicPLB14, CUJET3JHEP16, PHSDPRC16}
and Langevin dynamics~\cite{Akamatsu09, Das10, HFModelHee13, CaoPRC15, MCATHQPRC16}
to study the suppression and collective effect of the final heavy-flavor productions (having the charm or bottom quarks among these valence quarks)
such as D mesons ($D^{0}$, $D^{+}$, $D^{*+}$ and $D^{+}_{s}$~\cite{ShuangHP13, ShuangQM14})
and B mesons ($B^{0}$, $B^{+}$ and $B_{\rm s}$~\cite{BMesonCMS17}).
Comparing the theoretical calculations with available data,
it was realized~\cite{JFLPRL09, Das15, JFLCPL15, CTGUHybrid1} that the simultaneous description of $\raa$ and $\vtwo$                   
of open charmed meson at low and intermediate $\pt$ is sensitive to the temperature and energy dependence of the transport coefficients.
It is necessary to mention that, in order to improve the description of the measurements,
the Duke group~\cite{DukeBayesianPRC17}
develops a data-based hybrid model to extract the transport coefficient
by utilizing the Bayesian model-to-data analysis.
See Refs.~\cite{RalfSummary18, CaoCoefficient18, XuCoefficient18} for the recent review.

As mentioned, the Langevin approach is a very convenient
and widely used model, and it allows to establish, directly, a link between
the observables and transport coefficients, which can be extracted from the lattice QCD calculations.
However, the condition $m_{\rm Q}\gg gT$ may not always be justified,
in particular for charm quark with the medium temperature
close to its initial value,
resulting in the possible modification of the heavy meson $\raa$.
So, in this work,  we focus on the discussion related to the ``benefits and limitations for Boltzmann vs.
Langevin implementations of the heavy-flavor transport in an evolving medium''~\cite{RalfSummary18}.
Both the BTE and LTE will be employed to investigate the temperature
and energy dependence of the various transport coefficients,
as well as to study the charm quark transport behaviors in the QGP medium.

The paper is organized as follows.
In Sec.~\ref{sec:BoltLang} we summarize the employed Boltzmann and Langevin dynamics,
together with the comparison for the extracted transport coefficients including only the elastic interactions.
Sec.~\ref{sec:Method} is dedicated to the description of the hybrid model,
including the initial state configuration, the hydrodynamic expansion of the underlying medium,
heavy quark propagation and hadronization via fragmentation and ``heavy-light'' coalescence mechanisms.
Sec.~\ref{sec:Results_Col} shows the results obtained at parton and hadron levels with only the elastic processes,
while Sec.~\ref{sec:Results_ColRad} with both the elastic and inelastic contributions.
Sec.~\ref{sec:Summary} contains the summary and discussion.
\section{Boltzmann and Langevin Dynamics with only elastic processes}\label{sec:BoltLang}
\subsection{Linearized Boltzmann transport model}\label{subsec:Bolt}
The Boltzmann Transport Equation (BTE) reads
\begin{equation}\label{eq:BTE}
\frac{p_{\rm Q}}{E_{\rm Q}}\cdot \partial f_{\rm Q} = C[f_{\rm Q}]
\end{equation}
where, $p_{\rm Q}$, $E_{\rm Q}$ and $f_{\rm Q}$ are the HQ 4-momentum,
energy and distribution function, respectively.
$C[f_{\rm Q}]$ denotes the collision integral, including all the interaction mechanisms
between heavy quarks and the medium partons.
Equation~\ref{eq:BTE} can be linearized by ignoring the change of thermal parton distribution
in the medium due to the heavy quark propagation, and thus, $C[f_{\rm Q}]$
becomes a linear function of $f_{\rm Q}$.
Based on the Monte Carlo techniques, Eq.~\ref{eq:BTE} can be solved numerically by 
slicing the coordinate space into a 3-dimension grid,
and then the test particle method~\cite{WongPRC82} is used to sample $f_{\rm Q}$ in each cell.
The collision integral is solved by using the stochastic algorithm for evaluating the collision probability~\cite{XuPRC05, GrecoPLB13}.
In this work, we utilize only the linearized Boltzmann module in the Lido hybrid model~\cite{Lido18}
with all the default parameters, except the charm quark mass $m_{\rm c}=1.5~{\rm GeV}$.

In the local rest frame (LRF) of the cell,
the heavy quark transport is performed within a given time-step $\Delta t$.
Concerning a desired scattering process $l$, there are $n$ ($m$) incoming (outgoing) partons,
and the reaction probability $\Delta P_{l}$ is expressed as~\cite{Lido18}
\begin{equation}\label{eq:ReacProb}
\frac{\Delta P_{l}}{\Delta t} = \Gamma_{l}(E_{\rm Q},T,t) =
\frac{g}{\nu} \frac{(2\pi)^{3}\delta}{\delta f_{\rm Q}} \int d\Phi(n,m) \prod_{\{in\}}f_{i}\overline{{|M|}_{l}^{2}}
\end{equation}
where, $\Gamma_{l}(E_{\rm Q},T,t)$ is the relevant scattering rate;
$g$ is the spin-color degeneracy factor of the incoming medium partons;
$\nu$ is the statistical factor that corrects for double-counting when
there are identical particles in the initial/final state;
$f_{i}$ denotes the heavy quark ($i=Q$) and medium parton ($i=\bar{q},q,g$) density,
while the latter one follows the $\rm Maxwell-J{\ddot{u}}ttner$ distribution;
$\overline{{|M|}_{l}^{2}}$ is the initial state spin-color averaged scattering matrix element squared for two-body interactions,
which can be calculated via the perturbative QCD at leading-order~\cite{Combridge78}.
$d\Phi(n,m)$ in Eq.\ref{eq:ReacProb} is the $n+m$ body phase space integration,
\begin{equation}\label{eq:BodyPhSp}
d\Phi(n,m) = (2\pi)^{4} \delta^{(4)}(p_{\rm in} - p_{\rm out}) \prod_{\{in,out\}} \frac{d^{3}\vec{p}_{i}}{2E_{i}(2\pi)^{3}}
\end{equation}
where, $p_{\rm in}$ ($p_{\rm out}$) indicates the total 4-momentum of all the incoming (outgoing)
partons for a given $2\rightarrow2$ scattering process $l$.
Within the time interval $\Delta t$, the total reaction probability $\Delta P_{total}$ is given by
\begin{equation}\label{eq:TotalProb}
\Delta P_{total} = \sum_{l}(\Delta P_{l}) = \sum_{l}(\Gamma_{l} \cdot \Delta t).
\end{equation}

It was argued~\cite{Moore04} that the interactions between HQs and the medium partons can be encoded into the drag and momentum diffusion coefficients:
\begin{equation}
\begin{aligned}\label{eq:BTECoef}
&\eta_{\rm D} \equiv -\frac{d<p>}{dt}/<p> \\
&\kappa_{\rm L} \equiv \frac{d<(\Delta p_{\rm z})^{2}>}{dt} \\
&\kappa_{\rm T} \equiv \frac{1}{2} \frac{d<(\Delta p_{\rm T})^{2}>}{dt},
\end{aligned}
\end{equation}
which describes the average momentum/energy loss, momentum fluctuations in the direction that parallel
(i.e. longitudinal) and perpendicular (i.e. transverse) to the propagation, respectively.

\subsection{Langevin transport model}\label{subsec:Lang}
While traversing the Quark-Gluon Plasma (QGP),
HQ suffers frequent but soft momentum kicks from the medium partons,
therefore, HQ behaves the Brownian motion, which can be described by the Langevin Transport Equation (LTE)~\cite{TAMU13}
\begin{equation}
\begin{aligned}\label{eq:LTE_Col}
&\frac{dx^{\rm i}}{dt}=\frac{p^{\rm i}}{E^{\rm i}}
& \\
&\frac{dp^{\rm i}}{dt}=F^{\rm i}_{\rm Drag} + F^{\rm i}_{\rm Diff}
\end{aligned}
\end{equation}

The deterministic drag force reads
\begin{equation}\label{eq:DragForce}
F^{\rm i}_{\rm Drag}=-\eta_{\rm D}(\vec{p},T) \cdot p^{\rm i},
\end{equation}
where $\eta_{\rm D}(\vec{p},T)$ is the drag coefficient.

The stochastic force which acts on the HQ is expressed as
\begin{equation}\label{eq:ThermalForce}
F^{\rm i}_{\rm Diff}=\frac{1}{\sqrt{dt}}C^{\rm ij}(t,\vec{p}+\xi d\vec{p},T)\rho^{j}
\end{equation}
with the Gaussian noise $\rho^{j}$ follows a normal distribution
\begin{equation}\label{eq:LTEnoise1}
P(\vec{\rho})=(\frac{1}{2\pi})^{3/2} exp\left \{-\frac{\vec{\rho}^2}{2}\right \},
\end{equation}
resulting in $<\rho^{i}>_{\rho}=0$ and $<\rho^{i}\rho^{j}>_{\rho}=\delta^{ij}$.
Therefore, there is no correlation for the random force between two different time scales $<F^{\rm i}_{\rm Diff}(t)F^{\rm j}_{\rm Diff}(t^{\prime})>_{\rho} \equiv C^{\rm ik}C^{\rm kj}\delta(t-t^{\prime})$,
indicating the uncorrelated random momentum kicks from the medium partons.
During the numerical implementation, as shown in Eq.~\ref{eq:ThermalForce},
the stochastic process depends on the specific choice of the momentum argument of
the covariance matrix, $C^{\rm ij}(t,\vec{p}+\xi d\vec{p},T)$,
via a parameter $\xi\in[0,1]$.
Typically, $\xi=0$ for pre-point Ito, $\xi=1/2$ for mid-point and $\xi=1$ for post-point
discretization scheme of the stochastic integral.
Finally, $C^{\rm ij}$ can be represented
in terms of the longitudinal ($\kappa_{\rm L}$) and transverse momentum diffusion coefficients ($\kappa_{\rm T}$)~\cite{BeraudoNPA09}, i.e.
\begin{equation}\label{eq:LTEtensor}
C^{\rm ij}(\vec{p},T) \equiv \sqrt{\kappa_{\rm L}(\vec{p},T)}\hat{p}^{i}\hat{p}^{j} + \sqrt{\kappa_{\rm T}(\vec{p},T)}(\delta^{ij}-\hat{p}^{i}\hat{p}^{j}),
\end{equation}
therefore, the relation between $\eta_{\rm D}$, $\kappa_{\rm L}$ and $\kappa_{\rm T}$ is given by
\begin{equation}
\begin{aligned}\label{eq:LTEdragdiffusion}
\eta_{\rm D}=&\frac{\kappa_{\rm L}}{2TE} + (\xi-1)\frac{1}{2p}\frac{\partial\kappa_{\rm L}}{\partial p} + \\
&\frac{d-1}{2p^{2}}  \left[\xi(\sqrt{\kappa_{\rm T}}+\sqrt{\kappa_{\rm L}})^{2} - (3\xi-1)\kappa_{\rm T} - (\xi+1)\kappa_{\rm L} \right],
\end{aligned}
\end{equation}
where $d=3$ denotes the spatial dimension. As pointed, HQ diffusions are  conveniently encoded in the three coefficients
$\eta_{\rm D}$, $\kappa_{\rm L}$ and $\kappa_{\rm T}$. Note that Eq.~\ref{eq:LTEdragdiffusion} can be reduced to
\begin{equation}\label{eq:PostPoint1}
\eta_{\rm D}=\frac{\kappa_{\rm L}}{2TE} - \frac{d-1}{2p^{2}}(\sqrt{\kappa_{\rm T}}-\sqrt{\kappa_{\rm L}})^{2}
\end{equation}
with the post-point scheme, i.e. $\xi=1$.
Following our previous analysis~\cite{CTGUHybrid1, CTGUHybrid2},
 a ``minimum model'' by assuming a isotropic momentum dependence of the diffusion coefficient, $\kappa_{\rm L} = \kappa_{\rm T} \equiv \kappa$,
 is adopted in this work,
although it is just validated at $p=0$ and,
they not exactly the same at $p\ne 0$ region from the analytical calculations~\cite{POWLANG09}.
Eq.~\ref{eq:PostPoint1} is therefore further reduced to
\begin{equation}\label{eq:PostPoint2}
\eta_{\rm D}=\frac{\kappa}{2TE},
\end{equation}
which is the so-called dissipation-fluctuation relation (or Einstein relation) in the non-relativistic approximation.

\subsection{Boltzmann vs. Langevin}\label{subsec:BoltVSLang}
\begin{figure}[!htbp]
\begin{center}
\vspace{-1.0em}
\setlength{\abovecaptionskip}{-0.1mm}
\setlength{\belowcaptionskip}{-1.5em}
\includegraphics[width=.46\textwidth]{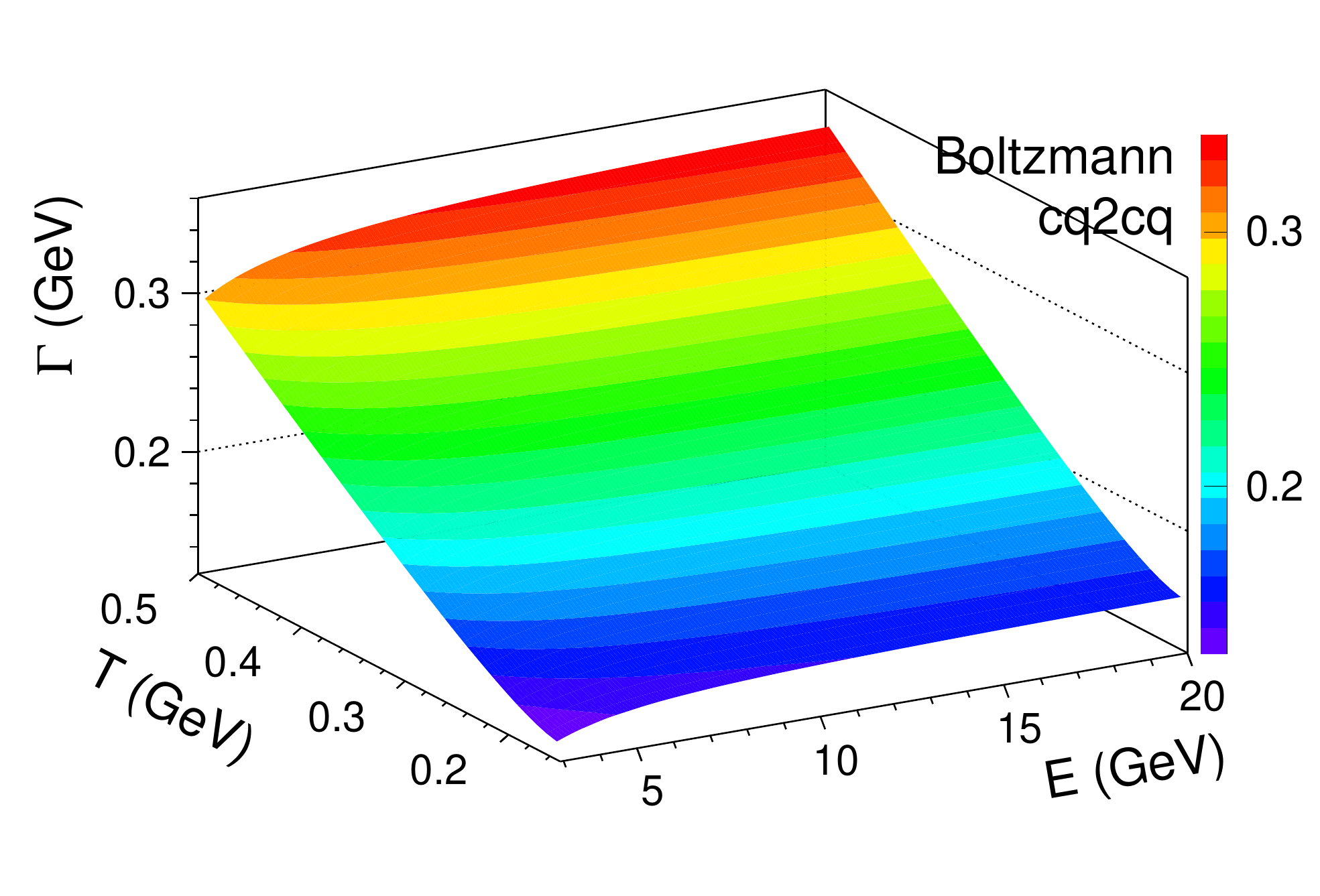}
\caption{(Color online) Scattering rate obtained in $c+q\rightarrow c+q$ via the Boltzmann model (Eq.~\ref{eq:ReacProb}).}
\label{fig:Gamma_c2qcq_vsET}
\end{center}
\end{figure}

In this sub-section, we mainly focus on the comparison of the transport coefficients
obtained via the Boltzmann and Langevin approaches with considering only the elastic scattering ($2\rightarrow2$) off the QGP constituents.
We show before that the scattering rate (Eq.~\ref{eq:ReacProb})
for $c+q\rightarrow c+q$ process in Fig.~\ref{fig:Gamma_c2qcq_vsET},
which is presented as a function of charm quark energy and the medium temperature.
It is found that the energy dependence is weak, while the temperature dependence is stronger.

\subsubsection{Boltzmann vs. Langevin: spatial diffusion coefficient}\label{subsubsec:2PiTDs}
The spatial diffusion coefficient $D_{s}$~\cite{Moore04} scaled by the thermal wavelength $1/(2\pi T)$,
\begin{equation}\label{eq:SpatialDiff}
2\pi TD_{s}=\lim_{p\rightarrow0}\frac{2\pi T^2}{m_{\rm Q} \cdot
\eta_{\rm D}(p)} = \lim_{E\rightarrow m_{\rm Q}}\frac{2\pi T^2}{m_{\rm Q} \cdot \eta_{\rm D}(E)},
\end{equation}
is defined at $p\rightarrow 0$ limit, which can be obtained directly by substituting Eq.~\ref{eq:BTECoef} with the Boltzmann approach.
$2\pi TD_{s}$  is available from the Lattice QCD calculation,
moreover, it is found~\cite{CTGUHybrid2} that, according to a phenomenological fitting analysis with the Langevin approach,
model predictions based on $2\pi TD_{s}=7$  allow to reproduce all the measured $\pt$ dependence
of the nuclear modification factor at both RHIC and LHC energies.
Therefore, in the Langevin approach, the drag and the momentum diffusion coefficients
(Eq.~\ref{eq:PostPoint2}) can be obtained via Eq.~\ref{eq:SpatialDiff} by setting $2\pi TD_{s}=7$.
Note that, in this case, (1) the definition of spatial diffusion coefficient is extended
to larger momentum values. Similar strategy is adopted in Ref.~\cite{AYPRC09, CaoPRC15, XuPRC17};
and (2) the drag and momentum diffusion coefficients in Eq.~\ref{eq:PostPoint2} can be represented in terms of $2\pi TD_{s}$ as
\begin{equation}
\begin{aligned}\label{eq:LTECoef}
&\eta_{\rm D}=\frac{1}{2\pi TD_{s}} \cdot \frac{2\pi T^{2}}{E} \\
&\kappa=\frac{1}{2\pi TD_{s}} \cdot {4\pi T^{3}}.
\end{aligned}
\end{equation}

\begin{figure}[!htbp]
\begin{center}
\vspace{-1.0em}
\setlength{\abovecaptionskip}{-0.1mm}
\setlength{\belowcaptionskip}{-1.5em}
\includegraphics[width=.46\textwidth]{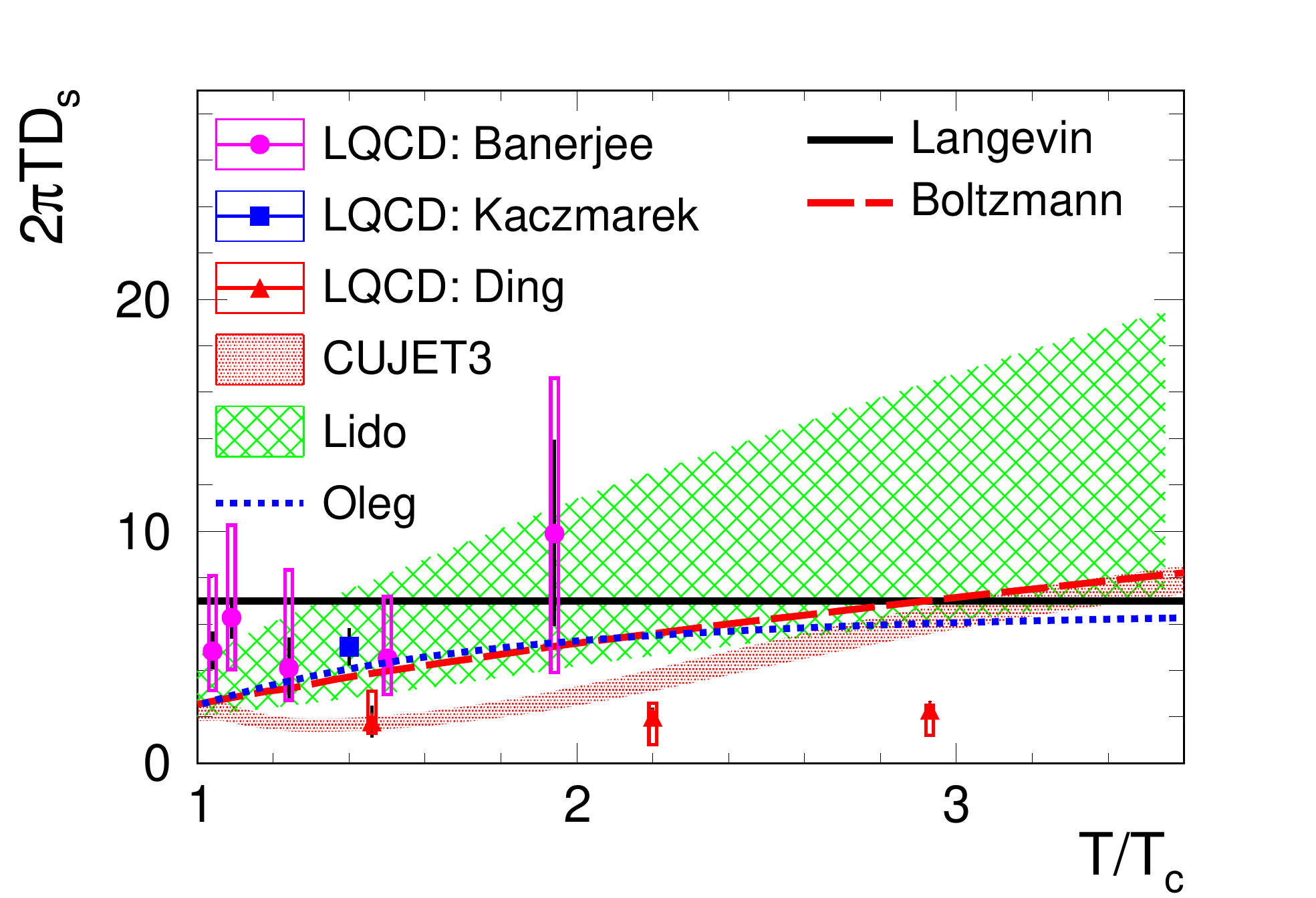}
\caption{(Color online) Spatial diffusion coefficient $2\pi TD_{s}$ of charm quarks ($m_{\rm c}=1.5$ GeV)
from lattice QCD calculations (pink circle~\cite{LQCDbanerjee12}, blue square~\cite{LQCDolaf14} and red triangle~\cite{LQCDding12} symbols)
at $p=0$. The CUJET3 (shadowing red region~\cite{CUJET3JHEP16}) and Lido model predictions (shadowing green region~\cite{Lido18}),
together with the results obtained from the Boltzmann (dashed red curve) and Langevin dynamics (solid black curve) are displayed for comparison.
Note that only the elastic scattering processes are considered with both the Boltzmann and Langevin approach.}
\label{fig:DsVsT}
\end{center}
\end{figure}
The temperature dependence of the spatial diffusion coefficient $2\pi TD_{s}$ is presented in Fig.~\ref{fig:DsVsT}.
The results from the Boltzmann (only $c+q\rightarrow c+q$ and $c+g\rightarrow c+g$) and Langevin (only collisional)
approaches are displayed as the dashed red and solid black curves, respectively. Lattice QCD and Ads/CFT calculations,
i.e. Banerjee (pink circles~\cite{LQCDbanerjee12}), Kaczmarek (blue square~\cite{LQCDolaf14}), Ding (red triangles~\cite{LQCDding12})
and Oleg (dotted blue curve~\cite{ADSCFTOleg}) are shown as well as for comparison.
Within the significant systematic uncertainties, both Boltzmann (dashed red curve) and Langevin predictions
(solid black curve) are consistent with the Banerjee and Oleg calculations.
Similar behavior can be found by comparing with the other model predictions,
such as the CUJET3 (red region)~\cite{CUJET3JHEP16} and Lido (green region)~\cite{Lido18}.

With Eq.~\ref{eq:SpatialDiff}, the thermalization time of charm quark, defined in $p\rightarrow 0$ limit~\cite{Moore04}, can be expressed as
\begin{equation}\label{eq:TauThermal}
\tau_{charm} \equiv \lim_{p\rightarrow0} {\eta_{\rm D}(p)}^{-1} =\frac{m_{charm}}{2\pi T^{2}_{c}} \cdot \frac{(2\pi TD_{s})}{(T/T_{c})^{2}},
\end{equation}
which is about 2.27 and 3.07 ${\rm fm/{\it c}}$ for Boltzmann and Langevin approach, respectively,
with $T=2T_{c}\approx330~{\rm MeV}$ and $m_{charm}=1.5~{\rm GeV}$.

\subsubsection{Boltzmann vs. Langevin: transport diffusion coefficients}\label{subsubsec:DiffCoe}
\begin{figure*}[!htbp]
\begin{center}
\vspace{-1.0em}
\setlength{\abovecaptionskip}{-0.1mm}
\setlength{\belowcaptionskip}{-1.5em}
\includegraphics[width=.32\textwidth]{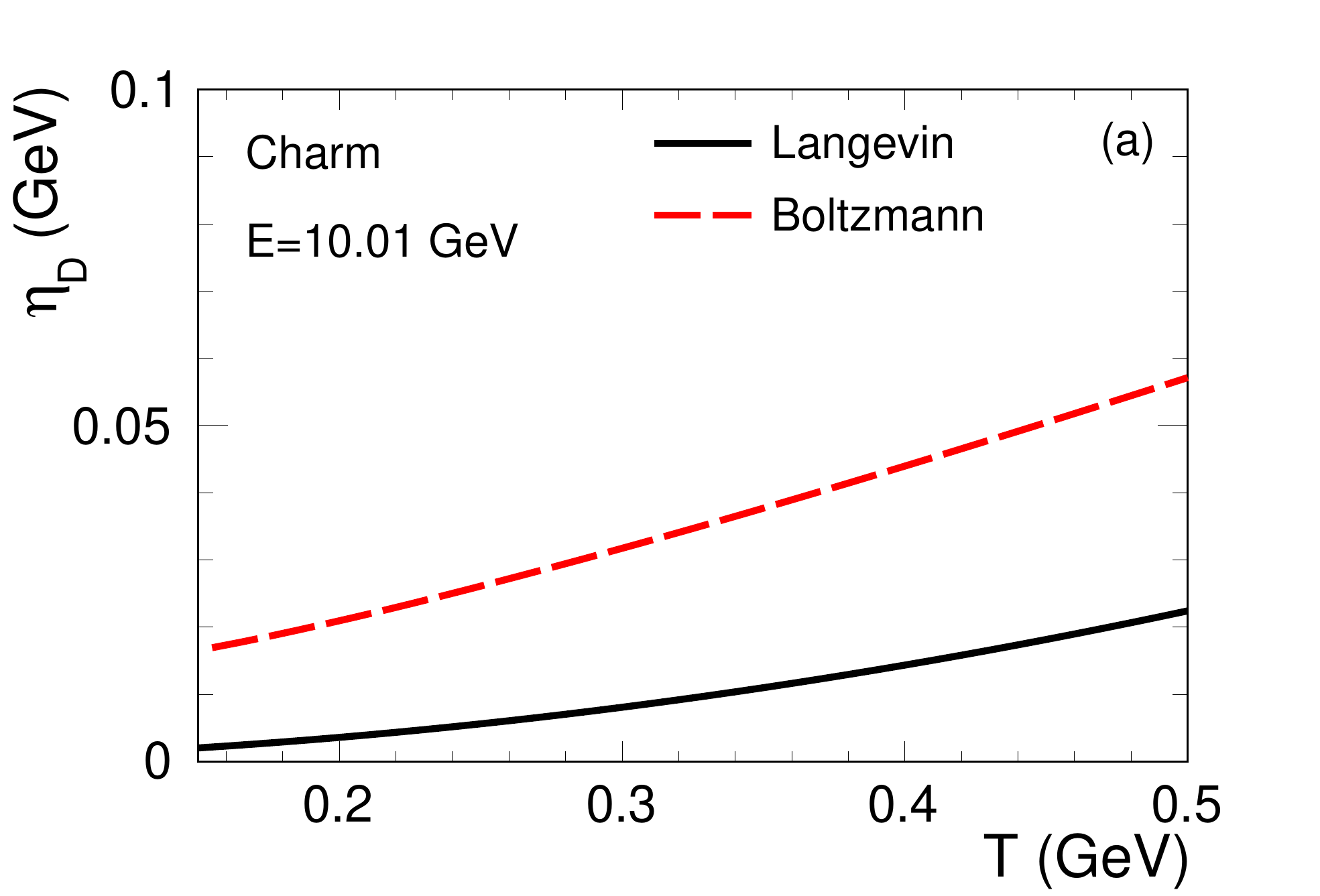}
\includegraphics[width=.32\textwidth]{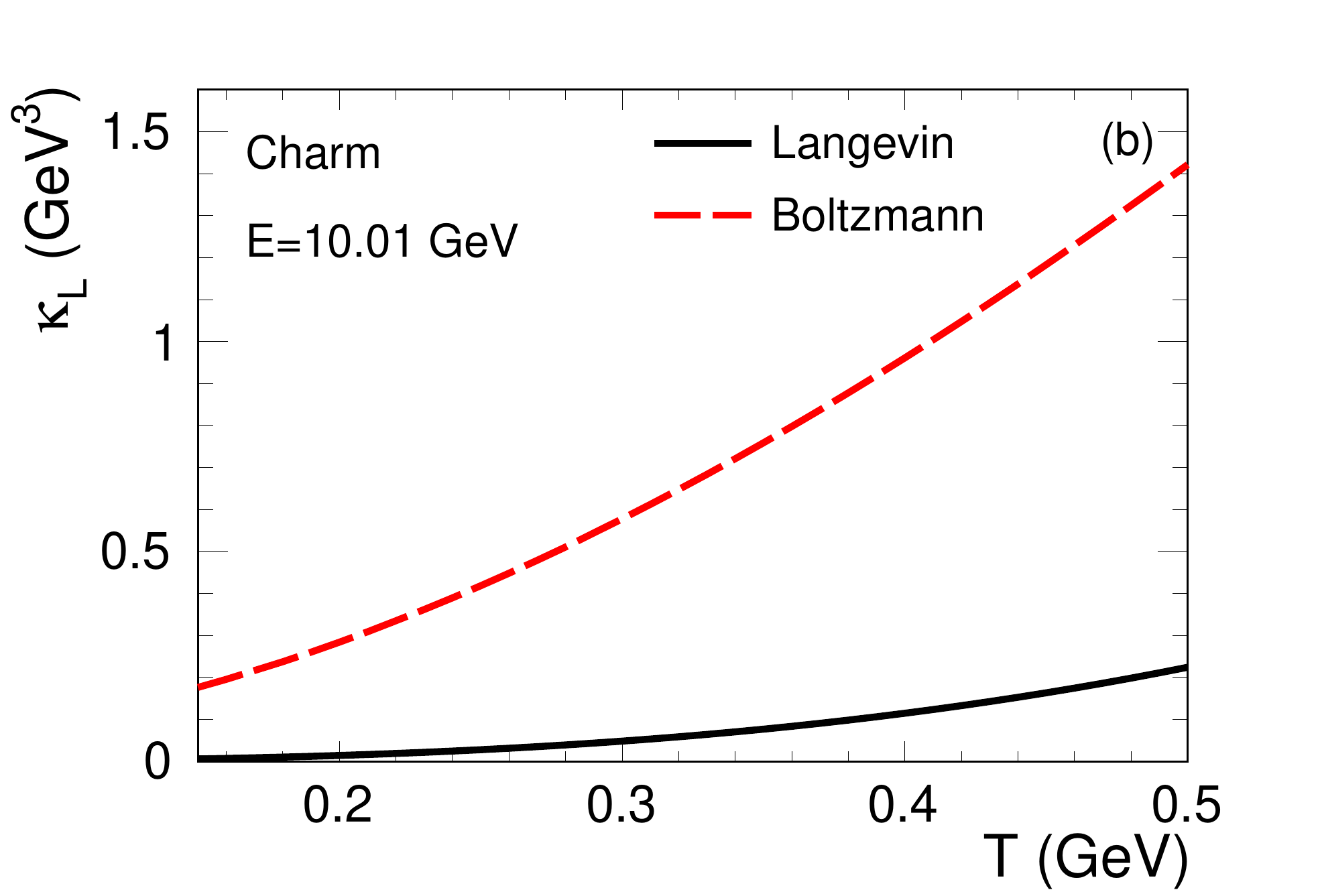}
\includegraphics[width=.32\textwidth]{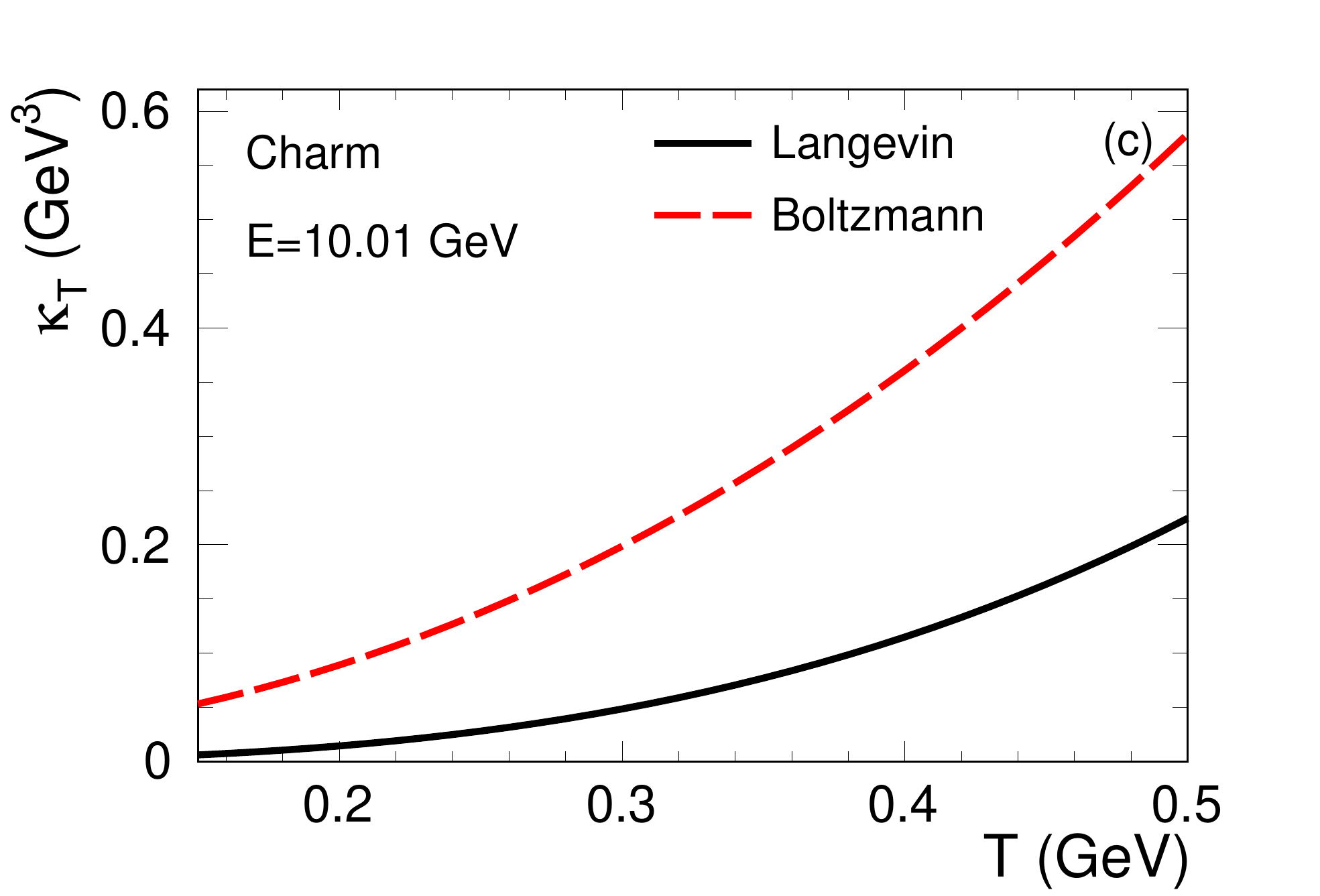}
\includegraphics[width=.32\textwidth]{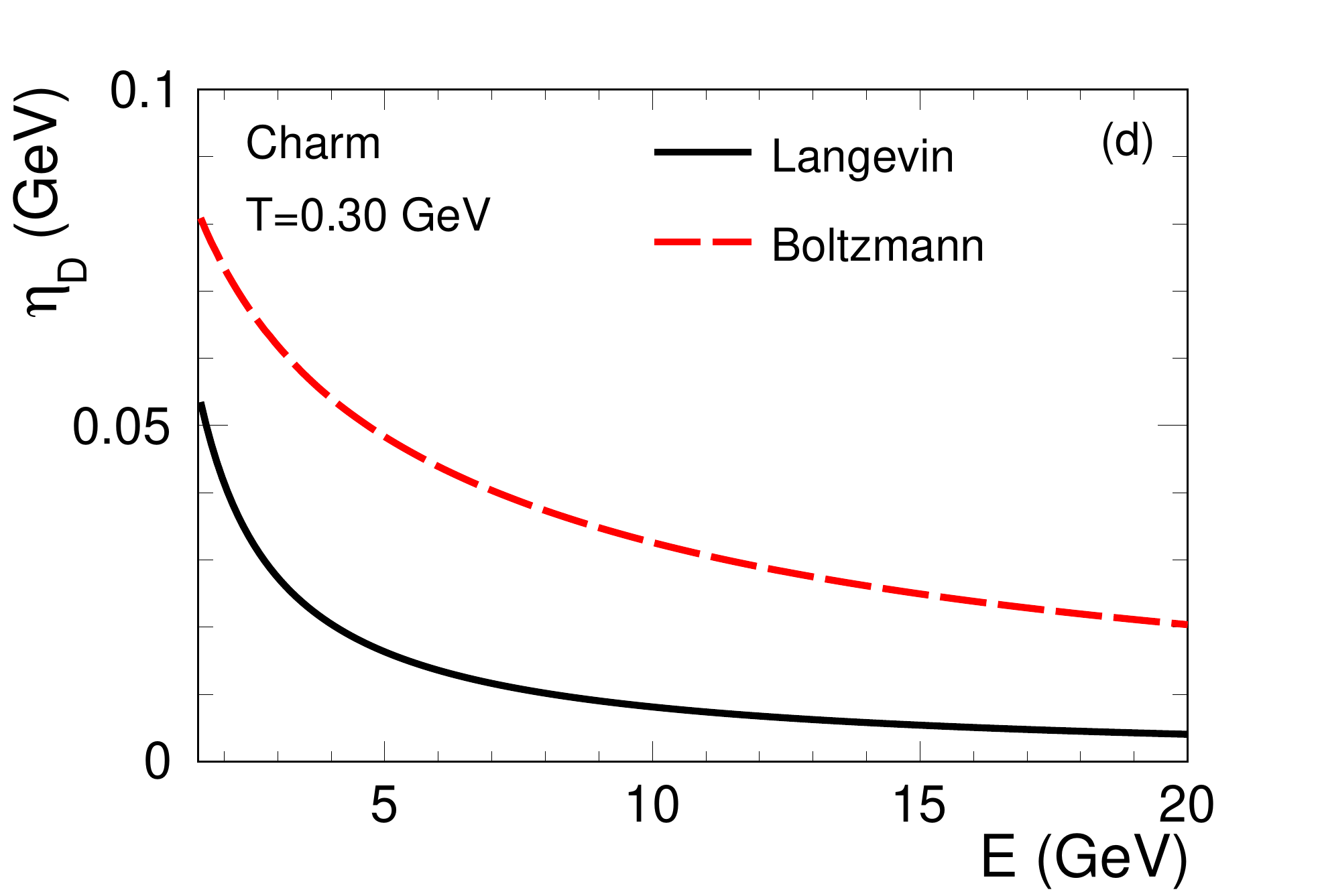}
\includegraphics[width=.32\textwidth]{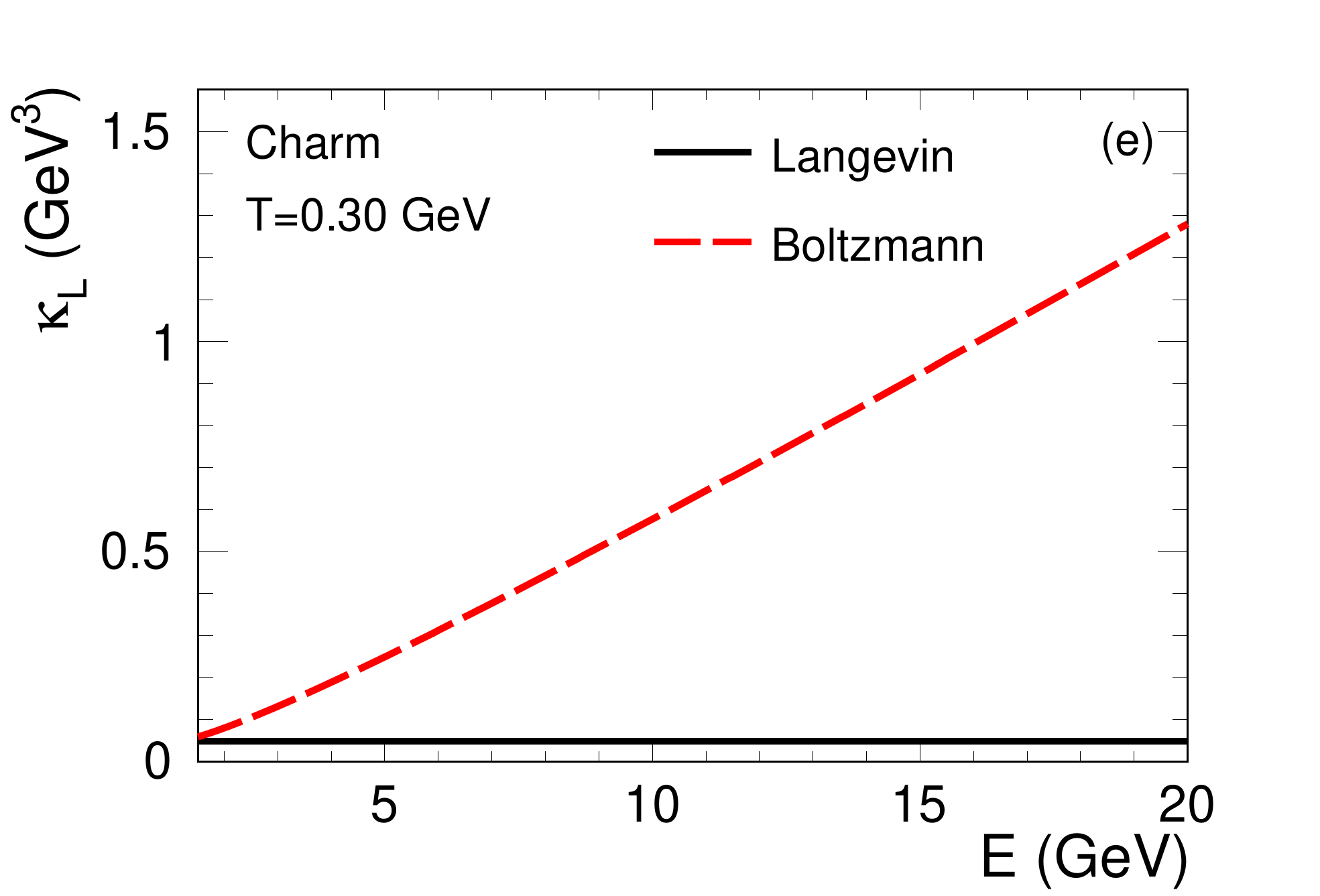}
\includegraphics[width=.32\textwidth]{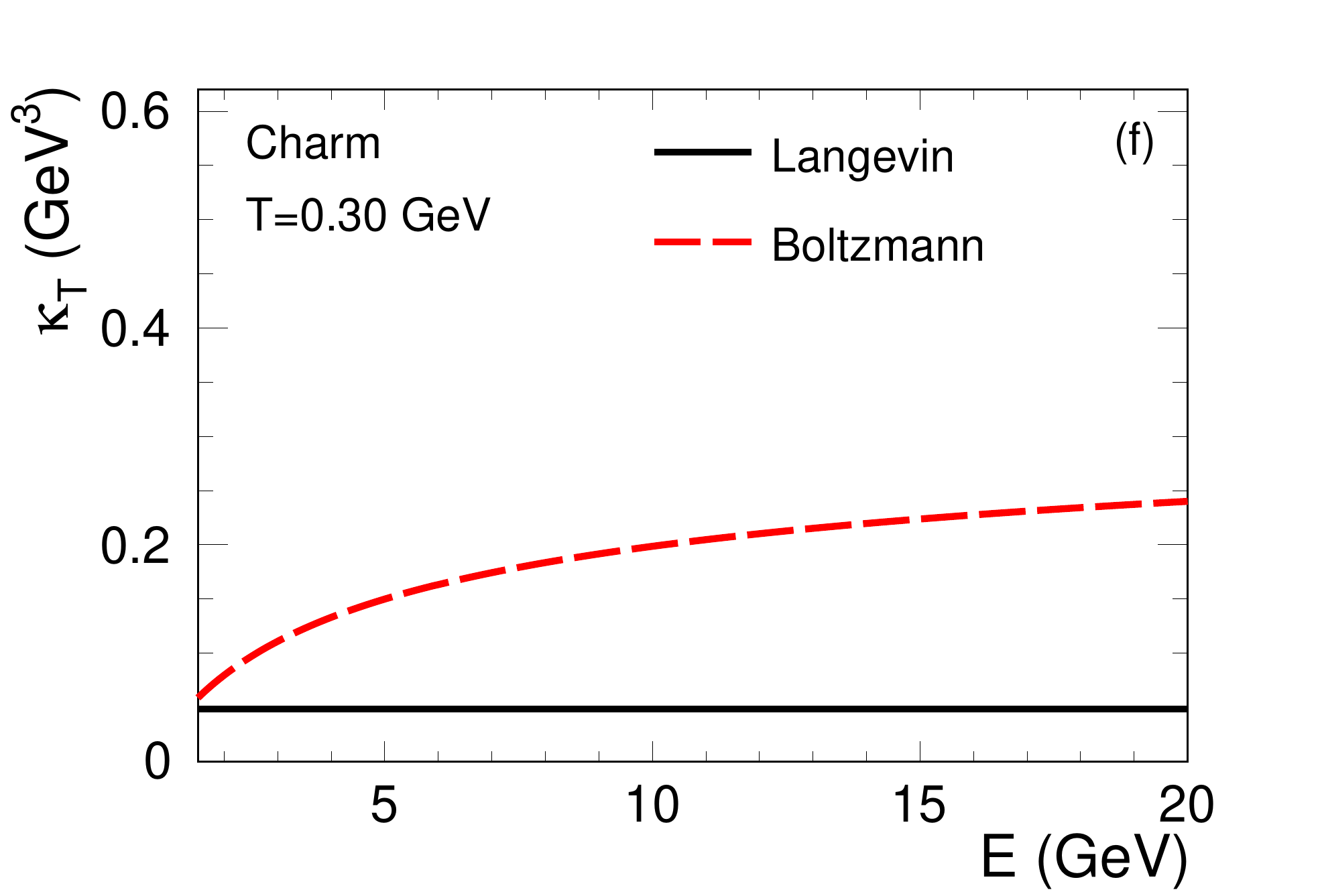}
\caption{(Color online) The drag coefficient $\eta_{\rm D}$ [(a), (d)], longitudinal $\kappa_{\rm L}$ [(b), (e)] and transverse momentum diffusion coefficients $\kappa_{\rm T}$ [(c), (f)]
with the Boltzmann (dashed red curve) and Langevin  model (solid black curve),
are shown at fix energy $E\approx10~{\rm GeV}$ (upper) and at fixed temperature  $T=0.3~{\rm GeV}$ (bottom).}
\label{fig:Coef_BoltLang_Col}
\end{center}
\end{figure*}

In Fig.~\ref{fig:Coef_BoltLang_Col}, the drag coefficient (left), longitudinal (middle) and transverse momentum diffusion coefficients (right),
are presented as a function of the medium temperature (upper) at a given energy $E\approx10~{\rm GeV}$,
and as a function of the charm quark energy (bottom) at fixed temperature $T=0.3~{\rm GeV}$.
The results obtained with the Boltzmann (Eq.~\ref{eq:BTECoef}) and Langevin model (Eq.~\ref{eq:LTECoef})
are shown as dashed red and solid black curves, respectively, in each panel.

Concerning the drag coefficient $\eta_{\rm D}$ [(a), (d)], the two models show an increasing temperature dependence and a decreasing behavior for the energy. The results with the Boltzmann approach (dashed red curve) is systematically larger than that with the Langevin approach (solid black curve).
Both the longitudinal $\kappa_{\rm L}$  [(b), (e)] and transverse momentum diffusion coefficient $\kappa_{\rm T}$  [(c), (f)] increases strongly with
increasing the energy and temperature via the Boltzmann approach,
while they change slowly, as expected (Eq.~\ref{eq:LTECoef}), via the Langevin approach.

\section{Methodology}\label{sec:Method}
In the previous analysis~\cite{CTGUHybrid1}, we construct a theoretical framework to study the charm quark propagation in ultrarelativistic heavy-ion collisions. The general modules of the hybrid model are discussed in the following.
\subsection{Initial state configuration}\label{subsec:IC}
The initialization of the heavy quark pairs is performed in the spatial and momentum space, respectively.
In the transverse direction, the initial spatial distribution is sampled
according to the initial binary collision density that is modeled by a Glauber-based approach~\cite{iEBE},
while in the longitudinal direction, it is described by a data-inspired phenomenological function~\cite{CTGUHybrid1}.
The initial momentum distribution of $c/\bar{c}$ is predicted by the FONLL calculations~\cite{FONLL98, FONLL01, FONLL12},
assuming a back-to-back azimuthal correlation between $c$ and $\bar{c}$ ($|\Delta\phi^{c\bar{c}}|=\pi$).
Fo nucleus-nucleus collisions, e.g. Pb--Pb, the nuclear modification of the parton distribution functions
(nPDFs) is taken into account by utilizing the EPS09 NLO parametrization approach~\cite{EPS09}.

The above initial state configuration allows providing the relevant
entropy density distribution, which will be taken as the input of the
subsequent hydrodynamical evolution.
All the parameters in this procedure are tuned by the model-to-data comparison~\cite{CTGUHybrid1}.
\subsection{Hydrodynamic description}\label{subsec:hydro}
The underlying medium evolution is modeled by
a 3+1D relativistic viscous hydrodynamics, vHLLE~{\cite{vhlle}},
with the initial time scale $\tau_{0}=0.6~{\rm fm}/{\it c}$, shear viscosity ${\eta/s=1/(4\pi)}$
and critical temperature $T_{c}=165~{\rm MeV}$ in both Au--Au and Pb--Pb collisions.
Note that the hydrodynamic simulation provides the space-time evolution of the temperature and the flow velocity field, which will be used in the HQ
Boltzmann and Langevin dynamics.

The QGP medium expands and cools down, and the (local) temperature
drops below the critical one $T_{c}$, resulting in the transition from the QGP phase to hadrons gas, namely hadronization.
After the transition, the hadron gas can in principle continue to interact inelastically until the chemical freeze-out, subsequently, the hadronic system
continues to expand and interact elastically until the kinetic freeze-out.
In this work, we neglect the chemical freeze-out procedure and consider, only, the kinetic freeze-out (or freeze-out since now)
occurs at $T_{c}=165~{\rm MeV}$.
An  instantaneous approach across a hypersurface of constant temperature,
namely isothermal freeze-out, is utilized and modeled by a widely used approach, Cornelius~\cite{cornelius}.

\subsection{Heavy quark propagation in medium}\label{subsec:HQDiffu}
We refer to Ref.~\cite{CTGUHybrid1} for the detailed discussion about the
numerical framework of charm quark Langevin evolution,
which is coupled with the expanding hydrodynamic medium.
For the Boltzmann case, it is quite similar except the procedure
to update the charm quark momentum in a discrete time-step.
In the following, we show the general strategy for both cases:
\begin{enumerate}
\item[(1)] sample a given number of HQ pairs at the position and momentum ($x^{\mu}, p^{\mu}$), in the laboratory frame (LAB),
according to the previous initial phase space configurations ($\tau\sim0$);
\item[(2)] move all the HQs from $\tau\sim0$ to $\tau_{0}=0.6~{\rm fm/{\it c}}$ as free streaming particles,
and modify the positions $x^{\mu}$ correspondingly;
\item[(3)] search the fluid cell at the same position as HQ, $x^{\mu}$,
and extract its temperature $T$ and velocity $u^{\mu}$ from the hydrodynamic simulations;
then, boost the current HQ to the local rest frame (LRF) of the fluid cell and get the HQ momentum in this frame;
\item[(4)] make a discrete time-step $\Delta t=0.01~{\rm fm/{\it c}}$ for the HQ in order to update its momentum $p^{\mu}$
        \begin{itemize}                                                                                                               
        \item Boltzmann dynamics: for the current HQ with $p^{\mu}_{old}$,
        calculate its reaction probability $\Delta P_{l}$ for each possible
        scattering channel $l$ (Eq.~\ref{eq:ReacProb});
        the target channel is selected according to the relative reaction
        probabilities $\Delta P_{l}/\Delta P_{total}$ (Eq.~\ref{eq:TotalProb}),
        meanwhile, the 4-momentum $p^{\mu}_{new}$ of the heavy quark after
        the scattering can be obtained according to
        the relevant scattering kinematics;
         \item Langevin dynamics: fix the drag and momentum diffusion coefficient
         with the fluid cell temperature $T$ (Eq.~\ref{eq:LTECoef}), as well as
         the drag (Eq.~\ref{eq:DragForce}) and
         thermal force (Eq.~\ref{eq:ThermalForce});
         and then, modify the HQ momentum $p^{\mu}$ according to the
         Langevin transport equation (Eq.~\ref{eq:LTE_Col});
         \end{itemize}                                                                                                               
\item[(5)] update the HQ position after the time step $\Delta t$
\begin{equation}
        x(t+\Delta t)-x(t)=\frac{p(t)}{E_{p}(t)}{\Delta t} \nonumber
        \end{equation}
        with the $p^{\mu}$ obtained in the previous step,
        and then boost back the HQ to the LAB frame;
\item[(6)] repeat the above steps (3)-(5) when the local temperature $T\geqslant T_{c}$.
\end{enumerate}

\subsection{Heavy quark hadronization via fragmentation and coalescence}\label{subsec:hadronization}
The heavy quark will suffer the instantaneous hadronization procedure
via a ``dual" approach, including fragmentation and heavy-light coalescence mechanisms, when the local temperature drops below the critical one $T_{c}=165~{\rm MeV}$. In this work, we follow the previous analysis~\cite{CTGUHybrid1} and
use this ``dual" model for the final heavy-flavor meson productions. 

Concerning the universal fragmentation function,
the Braaten approach~\cite{FragBraaten93} is employed in this work.
Due to the limitation of the measurements, it is difficult separate the open charmed hadrons
produced in decays of each excited charmed hadrons.
Practically, the relevant contributions can be treadted together with the fragmentation,
by including their contributions in the fragmentation fraction of a particular open charmed hadron.
Finally, the fragmentation fractions for the various hadron species are given by~\cite{CTGUHybrid1}
$f(c\rightarrow D^{0})=0.566$, $f(c\rightarrow D^{+})=0.227$, $f(c\rightarrow D^{\ast +})=0.230$,
$f(c\rightarrow D_{s}^{+})=0.081$ and $f(c\rightarrow \Lambda_{c})=0.080$.
The open charmed hadrons listed above are all the species included in the fragmentation model,
and the higher state contributions are considered,
which is consistent with the heavy-light coalescence model (to-be discussed below; Eq.~\ref{eq:SigM}).

The momentum distributions of heavy-flavor mesons ($Q\bar{q}$) reads
\begin{equation}
\begin{aligned}\label{eq:MesonCoal}
\frac{dN_{\rm M}}{d^{3}\vec{p}_{\rm M}}=&g_{\rm M}\int d^{3}\vec{x}_{\rm Q}d^{3}\vec{p}_{\rm Q} d^{3}\vec{x}_{\rm\bar{q}}d^{3}\vec{p}_{\rm\bar{q}} f_{\rm Q}(\vec{x}_{\rm Q},\vec{p}_{\rm Q}) f_{\rm\bar{q}}(\vec{x}_{\rm\bar{q}},\vec{p}_{\rm\bar{q}}) \\
&\times {\overline W}_{\rm M}^{\rm (n)}(\vec{y}_{\rm M},\vec{k}_{\rm M}) \delta^{(3)}(\vec{p}_{\rm M}-\vec{p}_{\rm Q}-\vec{p}_{\rm\bar{q}})
\end{aligned}
\end{equation}
where, $g_{\rm M}$ is the spin-color degeneracy factor;
$f_{\rm Q}(\vec{x}_{\rm Q},\vec{p}_{\rm Q})$ is the phase-space distributions of heavy quark, which can be obtained after the HQ propagate through the underlying QGP medium;
$f_{\rm\bar{q}}(\vec{x}_{\rm\bar{q}},\vec{p}_{\rm\bar{q}})$ is the one for light anti-quark,
which follows the Boltzmann-J$\ddot{\rm u}$ttner distribution in the momentum space
and it is spatially distributed on the freeze-out hypersurface.
The coalescence probability for $Q\bar{q}$ combination to form the heavy-flavor meson in the $n^{th}$ excited state,
is quantified by the overlap integral of the Wigner function of the meson
and the $Q\bar{q}$ pair~\cite{NewCoal16},
\begin{equation}
\begin{aligned}\label{eq:InteWig}
&{\overline W}_{\rm M}^{\rm (n)}(\vec{y}_{\rm M},\vec{k}_{\rm M}) \\
&=\int \frac{d^{3}\vec{x}^{\;\prime}_{\rm Q}d^{3}\vec{p}^{\;\prime}_{\rm Q}}{(2\pi)^{3}} \frac{d^{3}\vec{x}^{\;\prime}_{\rm\bar{q}}d^{3}\vec{p}^{\;\prime}_{\rm\bar{q}}}{(2\pi)^{3}} W_{\rm Q}(\vec{x}^{\;\prime}_{\rm Q}, \vec{p}^{\;\prime}_{\rm Q}) W_{\rm\bar{q}}(\vec{x}^{\;\prime}_{\rm\bar{q}}, \vec{p}^{\;\prime}_{\rm\bar{q}}) \\
&\quad \times W_{\rm M}^{\rm (n)}(\vec{y}^{\;\prime}_{\rm M}, \vec{k}^{\;\prime}_{\rm M}) \\
&=\biggr[ \frac{1}{2}\biggr(\frac{\vec{y}^{\;2}_{\rm M}}{\sigma_{\rm M}^{2}}+\sigma_{\rm M}^{2}\vec{k}^{\;2}_{\rm M}\biggr)\biggr]^{n} exp\biggr\{-\frac{1}{2}\biggr(\frac{\vec{y}^{\;2}_{\rm M}}{\sigma_{\rm M}^{2}}+\sigma_{\rm M}^{2}\vec{k}^{\;2}_{\rm M}\biggr)\biggr\} \biggr/ n!
\end{aligned}
\end{equation}
where, $\vec{y}_{\rm M}=(\vec{x}_{\rm Q}-\vec{x}_{\rm\bar{q}})$ and $\vec{k}_{\rm M}=(m_{\rm\bar{q}}\vec{p}_{\rm Q}-m_{\rm Q}\vec{p}_{\rm\bar{q}})/(m_{\rm Q}+m_{\rm\bar{q}})$
are the relative coordinate and the relative momentum, respectively, in the center-of-mass frame of $Q\bar{q}$ pair;
$W_{\rm Q}(\vec{x}^{\;\prime}_{\rm Q}, \vec{p}^{\;\prime}_{\rm Q})$ and $W_{\rm\bar{q}}(\vec{x}^{\;\prime}_{\rm\bar{q}}, \vec{p}^{\;\prime}_{\rm\bar{q}})$
are, respectively, the Wigner functions of heavy quark and light anti-quark
with their centroids at $(\vec{x}_{\rm Q},\vec{p}_{\rm Q})$ and $(\vec{x}_{\rm\bar{q}},\vec{p}_{\rm\bar{q}})$,
and they are both defined by taking the relevant wave function to be a Gaussian wave packet~\cite{CoalOriginalDover91}.
$W_{\rm M}^{\rm (n)}(\vec{y}^{\;\prime}_{\rm M}, \vec{k}^{\;\prime}_{\rm M})$
denotes the Wigner function of heavy-flavor meson, which is based on the well-known harmonic oscillator~\cite{CoalOriginalDover91}.
The width parameter $\sigma_{\rm M}$ is expressed as~\cite{CTGUHybrid1}
\begin{eqnarray}\label{eq:SigM}
\sigma_{\rm M}^{2}~&&= \left\{ \begin{array}{ll}
\frac{2}{3} \frac{(e_{\rm Q}+e_{\rm\bar{q}})(m_{\rm Q}+m_{\rm\bar{q}})^{2}}{e_{\rm Q}m_{\rm\bar{q}}^{2} + e_{\rm\bar{q}}m_{\rm Q}^{2}} \cdot \langle r_{\rm M}^{2} \rangle & \textrm{\qquad (n=0)} \\
\\
\frac{2}{5} \frac{(e_{\rm Q}+e_{\rm\bar{q}})(m_{\rm Q}+m_{\rm\bar{q}})^{2}}{e_{\rm Q}m_{\rm\bar{q}}^{2} + e_{\rm\bar{q}}m_{\rm Q}^{2}} \cdot \langle r_{\rm M}^{2} \rangle & \textrm{\qquad (n=1)}
\end{array} \right.
\end{eqnarray}
where, $\langle r_{\rm M}^{2} \rangle \approx (0.9~{\rm fm})^{2}$
is the mean-square charge radius of D-meson;
$e_{\rm Q}$ and $e_{\rm\bar{q}}$ are the absolute values of the charge of heavy quark and light anti-quark, respectively;
the light (anti-)quark mass takes $m_{\rm u/\bar{u}}=m_{\rm d/\bar{d}}=300~{\rm MeV}$ and $m_{\rm s/\bar{s}}=475~{\rm MeV}$.
We consider the various heavy-flavor meson species up to their first excited states ($n\leqslant1$),
see the Tab.~II as shown in Ref.~\cite{CTGUHybrid1} for details.

\begin{figure}[!htbp]
\begin{center}
\vspace{-1.0em}
\setlength{\abovecaptionskip}{-0.1mm}
\setlength{\belowcaptionskip}{-1.5em}
\includegraphics[width=.46\textwidth]{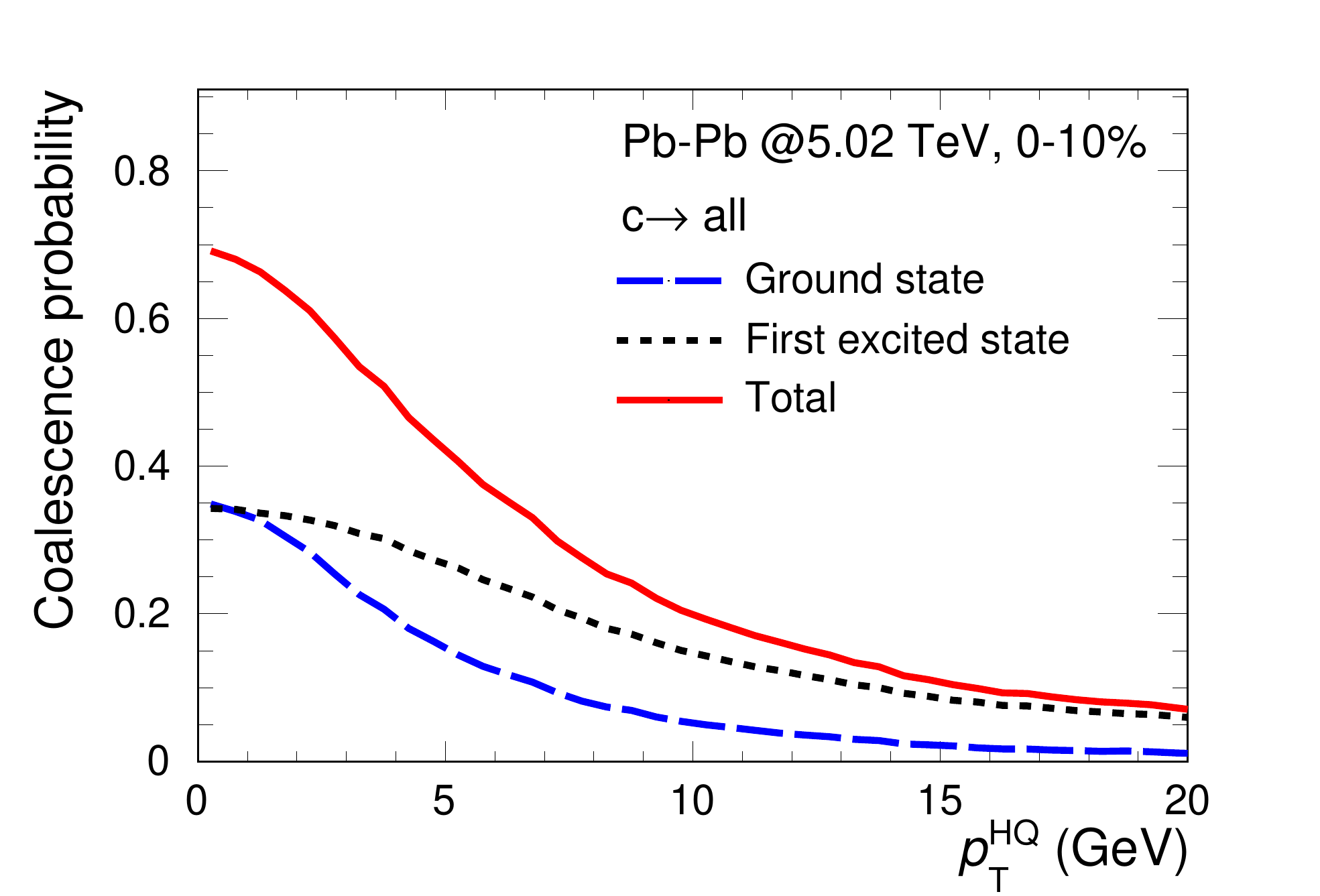}
\caption{(Color online) Comparison of the coalescence probability,
for $c\rightarrow$ D-meson in central ($0-10\%$) Pb--Pb collisions at $\snn=5.02~{\rm TeV}$,
contributed by the ground states (long dashed blue curve) and the first excited states (dashed black curve).
The combined results (solid red curve) are presented as well.}
\label{fig:CoalProbPbPb5020}
\end{center}
\end{figure}
In Fig.~\ref{fig:CoalProbPbPb5020} the coalescence probabilities obtained
in central ($0-10\%$) Pb--Pb collisions at $\snn=5.02~{\rm TeV}$,
are presented as a function of the charm quark transverse momentum.
The contributions of the ground states and the first excited states
are shown separately as the long dashed blue and short dashed black curves, respectively.
It is found that the coalescence into a ground state has maximum probability at $\pt^{\rm HQ}\sim0$,
and it decreases towards high $\pt$,
due to the difficulty to find a coalescence partner in this region.
The coalescence probability into the first excited states shows similar behavior.
The total coalescence probability is shown as a solid red curve,
which decreases from $\sim0.7$ at $\pt^{\rm HQ}\sim0$ to $0.2$ at $\pt^{\rm HQ}=10~{\rm GeV}$.
Moreover, the total coalescence probability is larger than $0.5$ in the range $\pt^{\rm HQ}\lesssim4~{\rm GeV}$,
reflecting its dominance in this region.

As displayed in Fig.~\ref{fig:CoalProbPbPb5020},
the hadronization of charm quark is divided into three channels:
fragmentation, coalescence to form D mesons at ground state and
at first excited state. During the implementation,
we generate a random number, using the Monte Carlo techniques, with flat distribution between zero and one,
and then compare it to the above three probabilities.
Finally, the target channel can be selected, and the momentum
of the relevant heavy-flavor meson will be obtained by assuming
the energy-momentum conservation in the $Q$ and $\bar{q}$ combination procedure.
See Ref.~\cite{CTGUHybrid1} for details.
\section{Results with considering only elastic processes}\label{sec:Results_Col}
\subsection{Momentum distribution inside a static medium}\label{subsec:Static}
In order to study the difference between the Boltzmann and Langevin
dynamics, in this sub-section, we focus on the time evolution
of the charm quark momentum distribution,
which is obtained inside a static medium with temperature fixed at $T=0.3~{\rm GeV}$,
as well as the momentum initialized at $p=10~{\rm GeV}$.

\begin{figure}[!htbp]
\begin{center}
\vspace{-1.0em}
\setlength{\abovecaptionskip}{-0.1mm}
\setlength{\belowcaptionskip}{-1.5em}
\includegraphics[width=.46\textwidth]{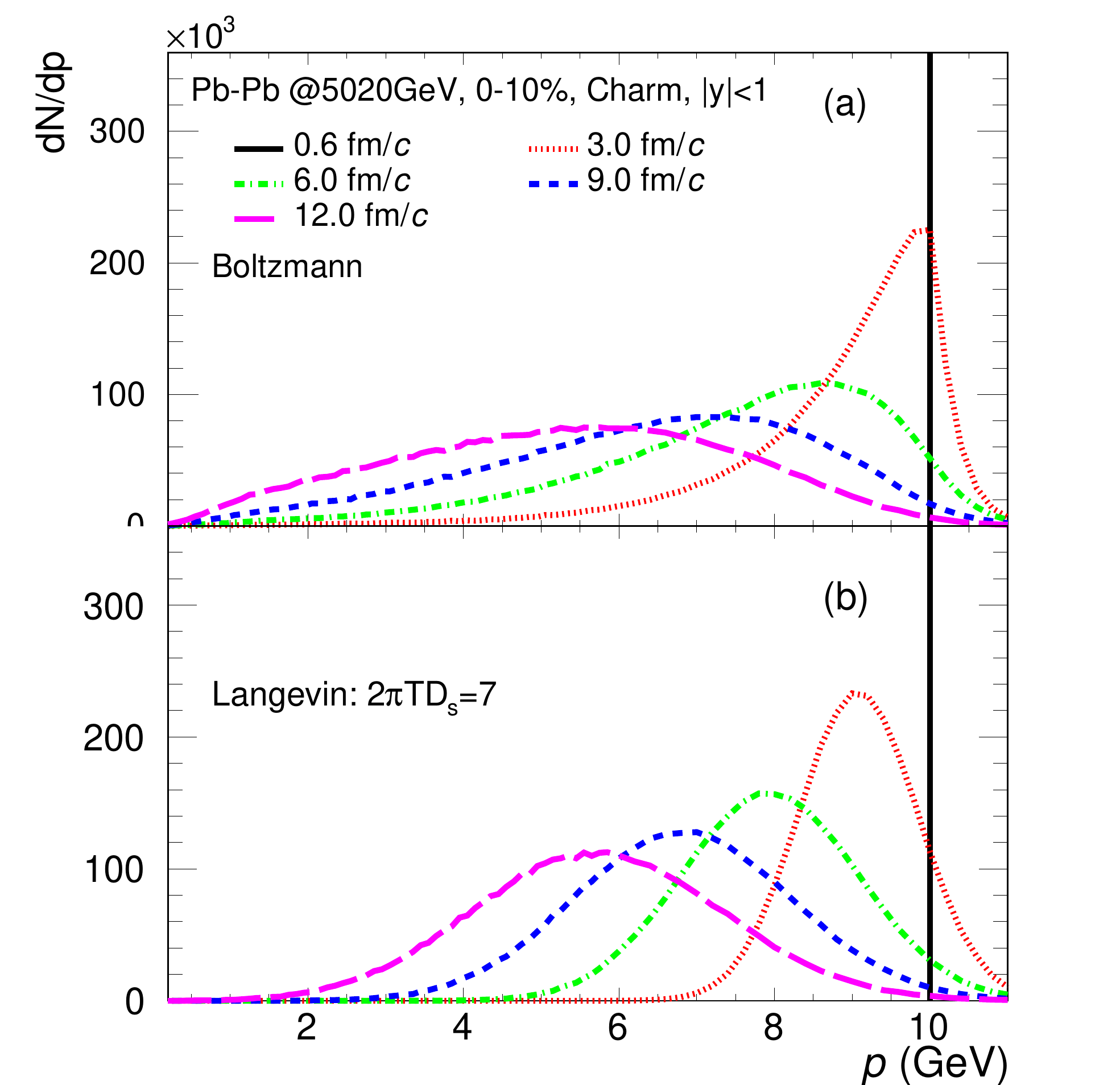}
\caption{(Color online) (a) Charm quark momentum distribution based on
the Boltzmann dynamics at different times during the
hydrodynamical evolution of the medium with a constant temperature $T=0.3~{\rm GeV}$ (see legend for details).
(b) similar as panel (a) but with the Langevin approach.
Only the collisional energy loss mechanism is considered.}
\label{fig:dNdP1_Static_Col}
\end{center}
\end{figure}

In Fig.~\ref{fig:dNdP1_Static_Col}, the charm quark momentum distribution $dN/dp$ based on the Boltzmann model [(a)],
is calculated at various times during the hydrodynamic evolution of the medium, showing as different styles.
At the starting time $\tau_{0}=0.6~{\rm fm/\it{c}}$ (solid black curve), as expected,
the initial $dN/dp$ behaves a delta distribution at $p=10~{\rm GeV}$.
During the evolution up to $\tau=12~{\rm fm/\it{c}}$,
$dN/dp$ is broadened comparing with the initial distribution,
meanwhile, the average momentum is shifted toward low $p$,
which is mainly induced by the drag force.
This is caused by the fact that the initial momentum spectrum of charm quark is much harder than that of medium parton,
and the multiple elastic scatterings are therefore dominated by the drag rather than the diffusion term~\cite{CTGUHybrid1}.
The results based on the Langevin approach [(b)] present
a different broadening behavior, which follows a Gaussian-like shape,
as expected in the construction (Eq.~\ref{eq:LTEnoise1}).
Similar results can be found in Ref.~\cite{HQBoltLang14}.
Comparing Boltzmann with Langevin calculations, it is observed that
the momentum broadening profile is stronger with the Boltzmann model,
since the scatterings with large momentum transfer are allowed in this approach, which are discarded with the Langevin approach.
Consequently, the relevant azimuthal angle distribution
with the Boltzmann model, is expected to show a stronger broadening behavior as compared to Langevin.
Note that, for both Boltzmann and Langevin dynamics,
$dN/dp (\tau=3~{\rm fm/\it{c}})$ (dotted red curve) is followed by a tail in the range $p>10~{\rm GeV}$,
where the interaction processes allow the charm quark to gain more energy respect to the lost term.

\subsection{Elastic energy loss inside a realistic medium}\label{subsec:Real}
\begin{figure}[!htbp]
\begin{center}
\vspace{-1.0em}
\setlength{\abovecaptionskip}{-0.1mm}
\setlength{\belowcaptionskip}{-1.5em}
\includegraphics[width=.46\textwidth]{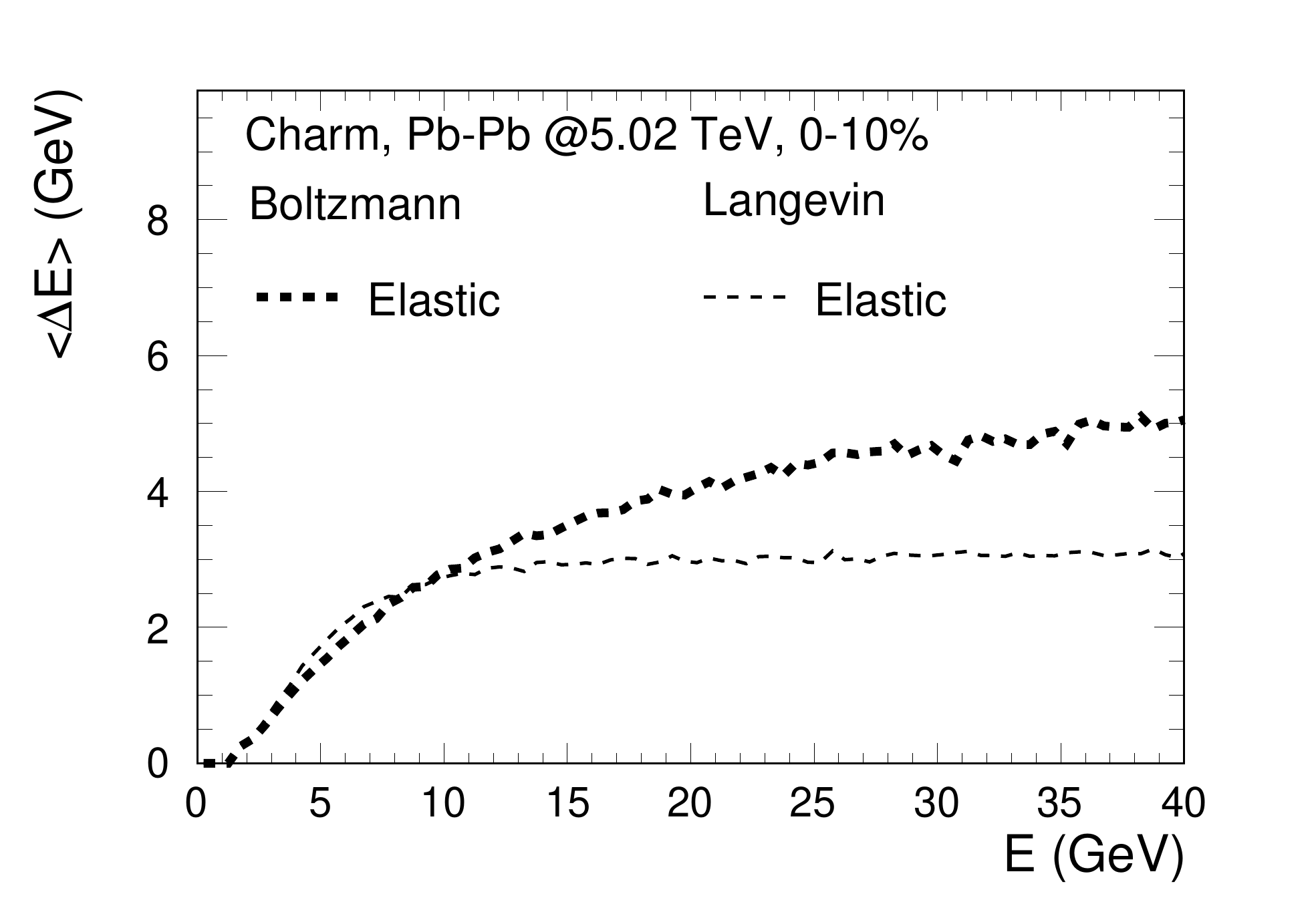}
\caption{Elastic energy loss of charm quarks obtained via Boltzmann approach (thick curve) and
Langevin approach with $2\pi TD_{s}=7$ (thin curve)
after the propagation through a realistic hydrodynamic medium.}
\label{fig:Eloss_Col}
\end{center}
\end{figure}

Figure~\ref{fig:Eloss_Col} shows the average in-medium energy loss of charm quark,
due to elastic scatterings, as a function of its initial energy.
The results with the Boltzmann and Langevin dynamics are presented as the thick and thin curve, respectively.
When comparing Boltzmann with Langevin results,
they are similar at low energy ($E\lesssim10~{\rm GeV}$) where the interactions with small momentum transfer are dominated,
while the former one is systematically larger than the latter one at higher energy,
resulting in a softer charm quark spectrum in this region.

\subsection{$\raa$ and $\vtwo$ for charm quarks}\label{subsec:PartonRAAV2}
\begin{figure}[!htbp]
\begin{center}
\vspace{-1.0em}
\setlength{\abovecaptionskip}{-0.1mm}
\setlength{\belowcaptionskip}{-1.5em}
\includegraphics[width=.46\textwidth]{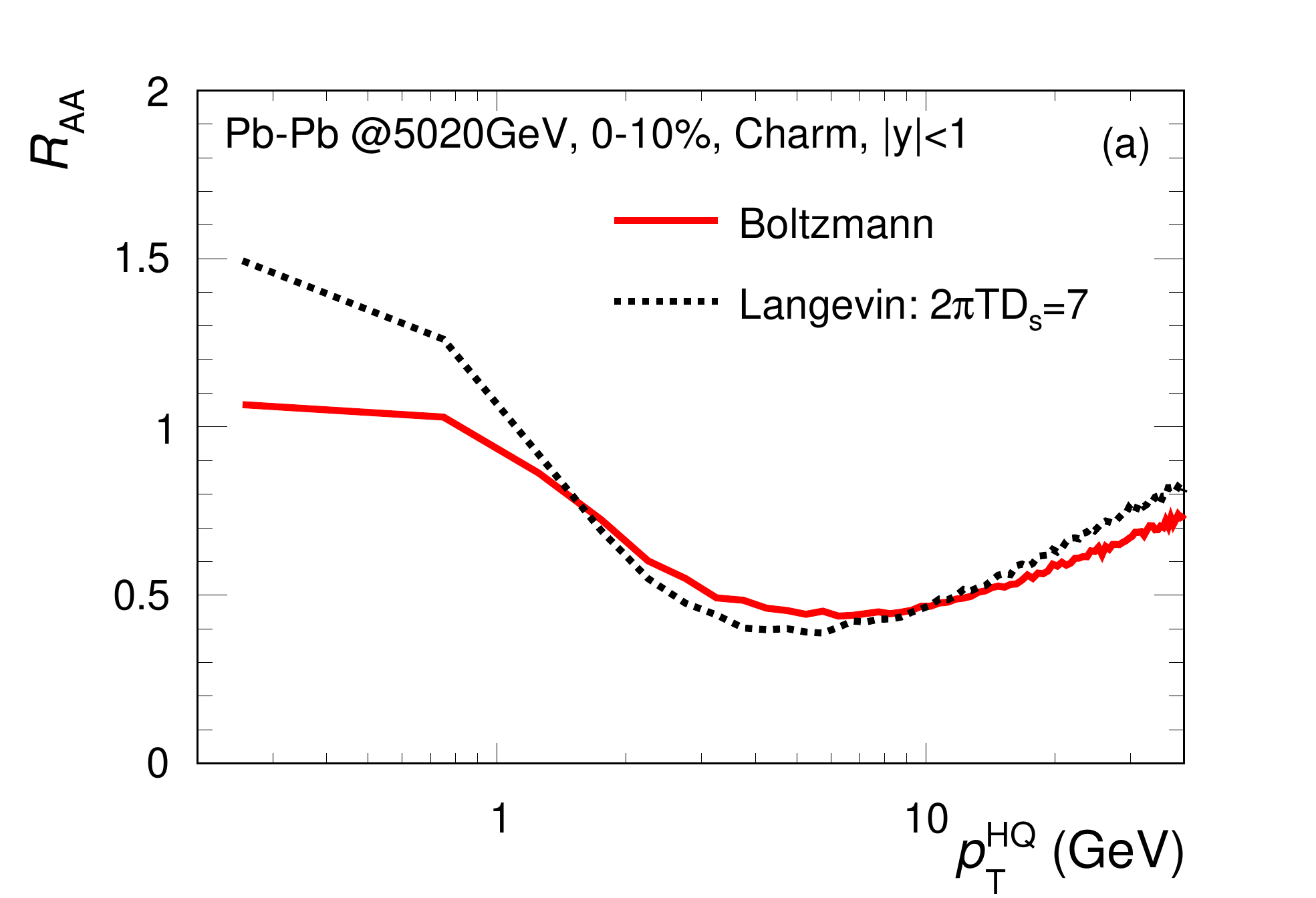}
\includegraphics[width=.46\textwidth]{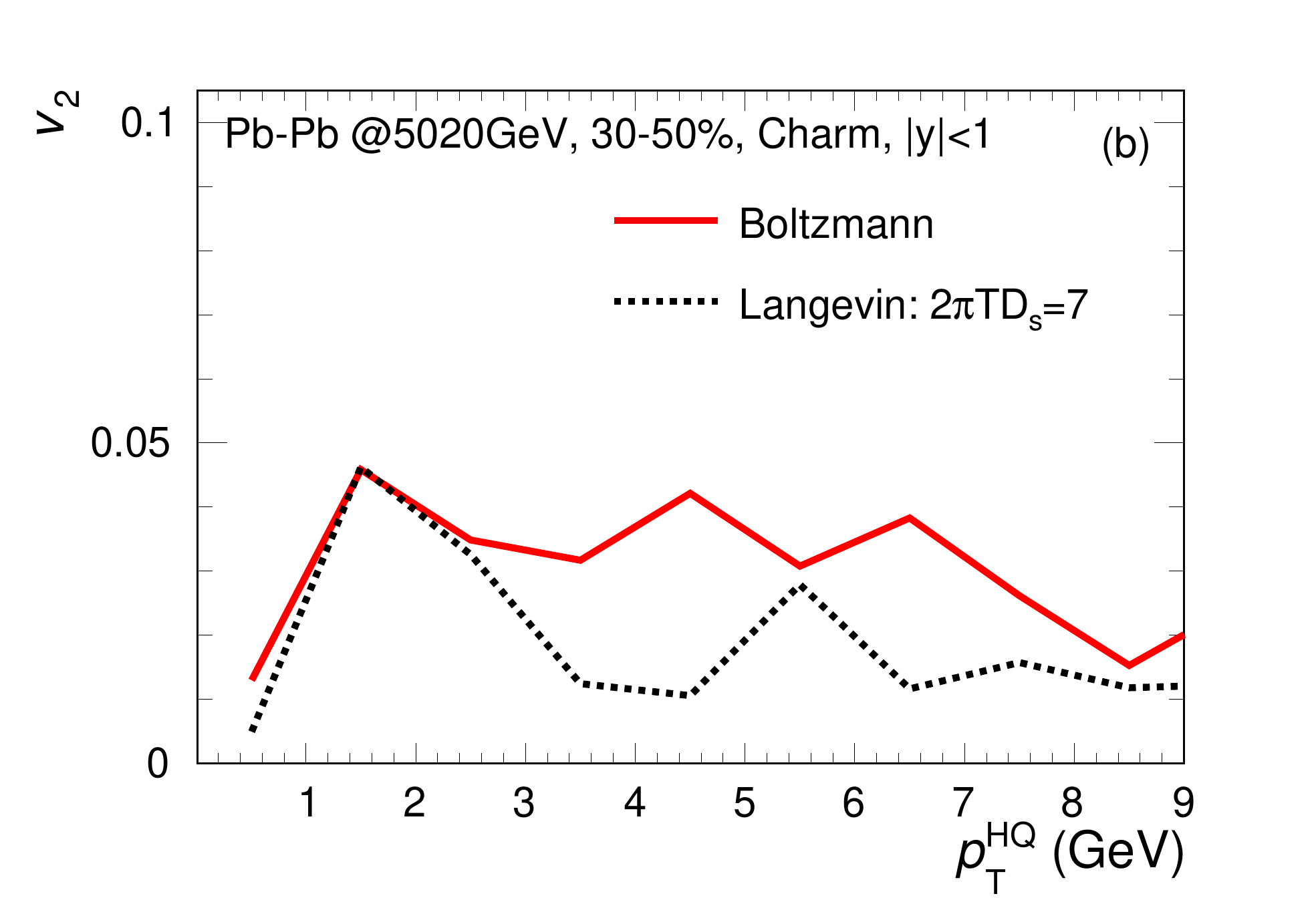}
\caption{(Color online) (a) Comparison of the charm quark $\raa$ obtained
with the Boltzmann (solid red curve) and Langevin approach (dashed black curve),
in central ($0-10\%$) Pb--Pb collisions at $\snn=5.02~{\rm TeV}$.
(b) Same as upper but for charm quark $\vtwo$
obtained in semicentral ($30-50\%$) Pb--Pb collisions at $\snn=5.02~{\rm TeV}$.
Only the collisional energy loss mechanism is considered.}
\label{fig:HQRAAV25020_Col}
\end{center}
\end{figure}

Figure~\ref{fig:HQRAAV25020_Col} shows the nuclear modification factor $\raa$ ($\vtwo$) of charm quark
obtained with the Boltzmann (solid red curves) and Langevin approach (dashed black curve) in central (semicentral) Pb--Pb collisions at $\snn=5.02~{\rm TeV}$.
It is observed that $\raa$, as displayed in the panel (a), is suppressed at high $\pt$ with
the Boltzmann approach as compared to the Langevin approach.
Therefore, charm quark loses more its initial energy while traversing
the medium in the Boltzmann dynamics,
which is consistent with results shown in Fig.~\ref{fig:Eloss_Col}.
The elliptic flow coefficient $\vtwo$, as presented in the panel (b),
with the Boltzmann approach is systematically larger as compared to the Langevin approach,
which means that the Boltzmann dynamics is more efficient in producing $\vtwo$.
Similar behavior is observed for different centrality classes and at different energies.
It is most probably due to fact that the drag coefficient is larger in Boltzmann,
resulting in a larger drag force acted on the charm quarks,
which is able to introduce more significant interactions with the QGP partons,
as well as to transfer more $\vtwo$ from the medium partons to the charm quarks.

\begin{figure}[!htbp]
\begin{center}
\vspace{-1.0em}
\setlength{\abovecaptionskip}{-0.1mm}
\setlength{\belowcaptionskip}{-1.5em}
\includegraphics[width=.46\textwidth]{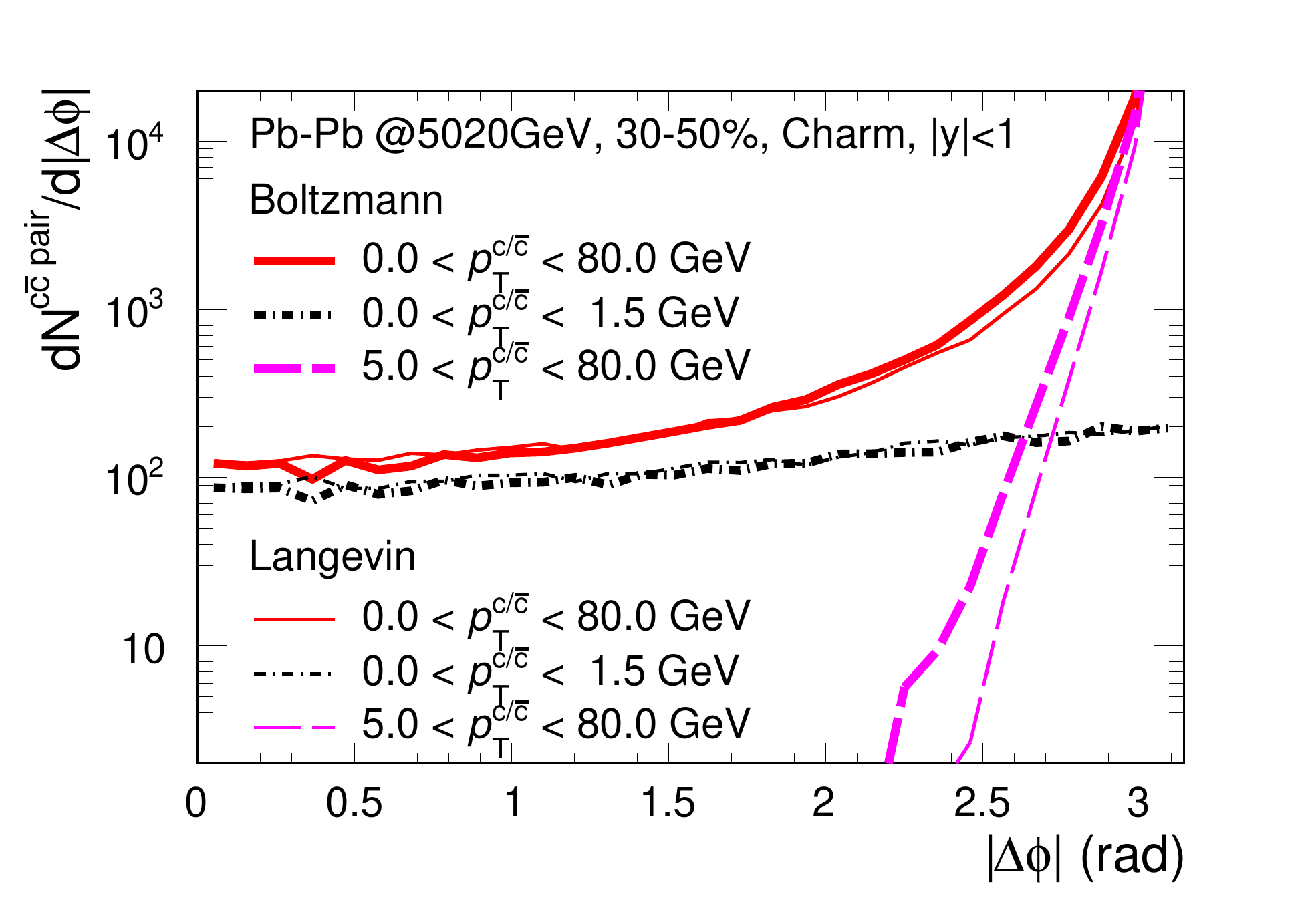}
\caption{(Color online) Comparison of the relative azimuthal angle between $c$ and $\bar{c}$ quarks
with the Boltzmann (thick curves) and Langevin approach (thin curves) in central ($0-10\%$) Pb--Pb collisions at $\snn=5.02~{\rm TeV}$.
The curves in different styles indicate the results within different $\pt$ intervals (see legend for details).
Only the collisional energy loss mechanism is considered.}
\label{fig:HQPhi5020_C0_Col}
\end{center}
\end{figure}

Concerning the relative azimuthal angle distribution,
the yields of the initially back-to-back generated $c\bar{c}$ pairs
can be described by a delta distribution at $|\Delta\phi|=\pi$.
After propagating through the medium, it is found that the above $|\Delta\phi|=\pi$ distribution is broadened
within different initial transverse momentum interval $\pt^{\rm c/\bar{c}}$,
as shown in different curves in Fig.~\ref{fig:HQPhi5020_C0_Col}.
With the Boltzmann approach (thick curves),
It is clear to see that there is an almost flag behavior with the lower initial transverse momentum
$\pt^{\rm c/\bar{c}}<1.5~{\rm GeV}$ (dotted black curve),
indicating the corresponding initially back-to-back properties are largely washed out
throughout the interactions with the surrounding medium constituents~\cite{CTGUHybrid1}.
Meanwhile, the broadening behavior tends to decrease with increasing $\pt^{\rm c/\bar{c}}$ (dashed pink curve).
Similar results can be found with the Langevin approach (thin curves).
Note that the nuclear (anti-)shadowing effect is not included.
Comparing Boltzmann with Langevin approach,
they are close within small $\pt^{\rm c/\bar{c}}$ region,
while the former one shows stronger broadening behavior at larger $\pt^{\rm c/\bar{c}}$.
This is because, as explained above, with the larger initial drag force,
the interactions in the Bolatzmann model are stronger and more powerful
to pull the $c\bar{c}$ pairs from high momentum to low momentum.

\subsection{$\raa$ and $\vtwo$ for heavy-flavor mesons}\label{subsec:MesonRAAV2}
\begin{figure}[!htbp]
\begin{center}
\vspace{-1.0em}
\setlength{\abovecaptionskip}{-0.1mm}
\setlength{\belowcaptionskip}{-1.5em}
\includegraphics[width=.46\textwidth]{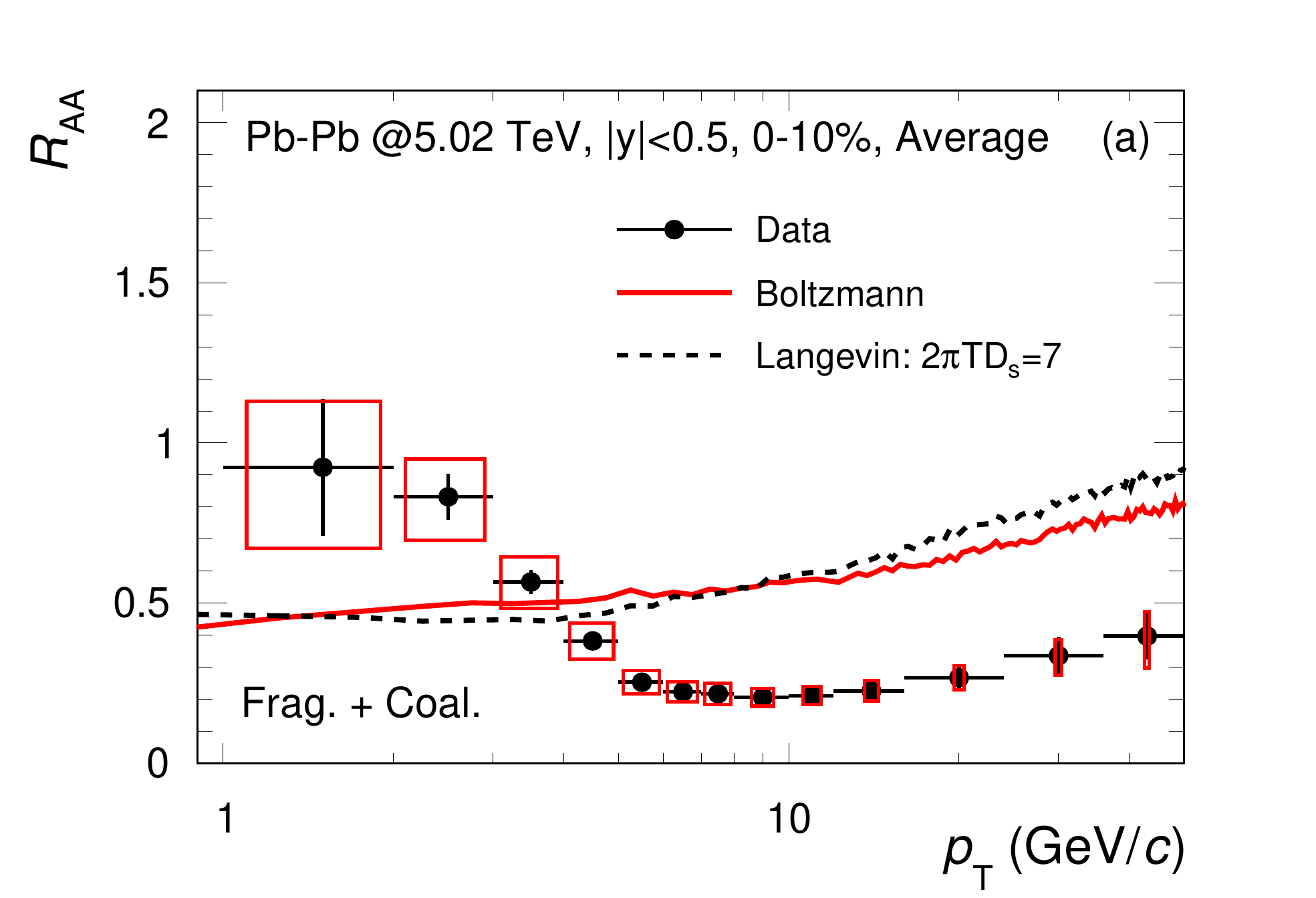}
\includegraphics[width=.46\textwidth]{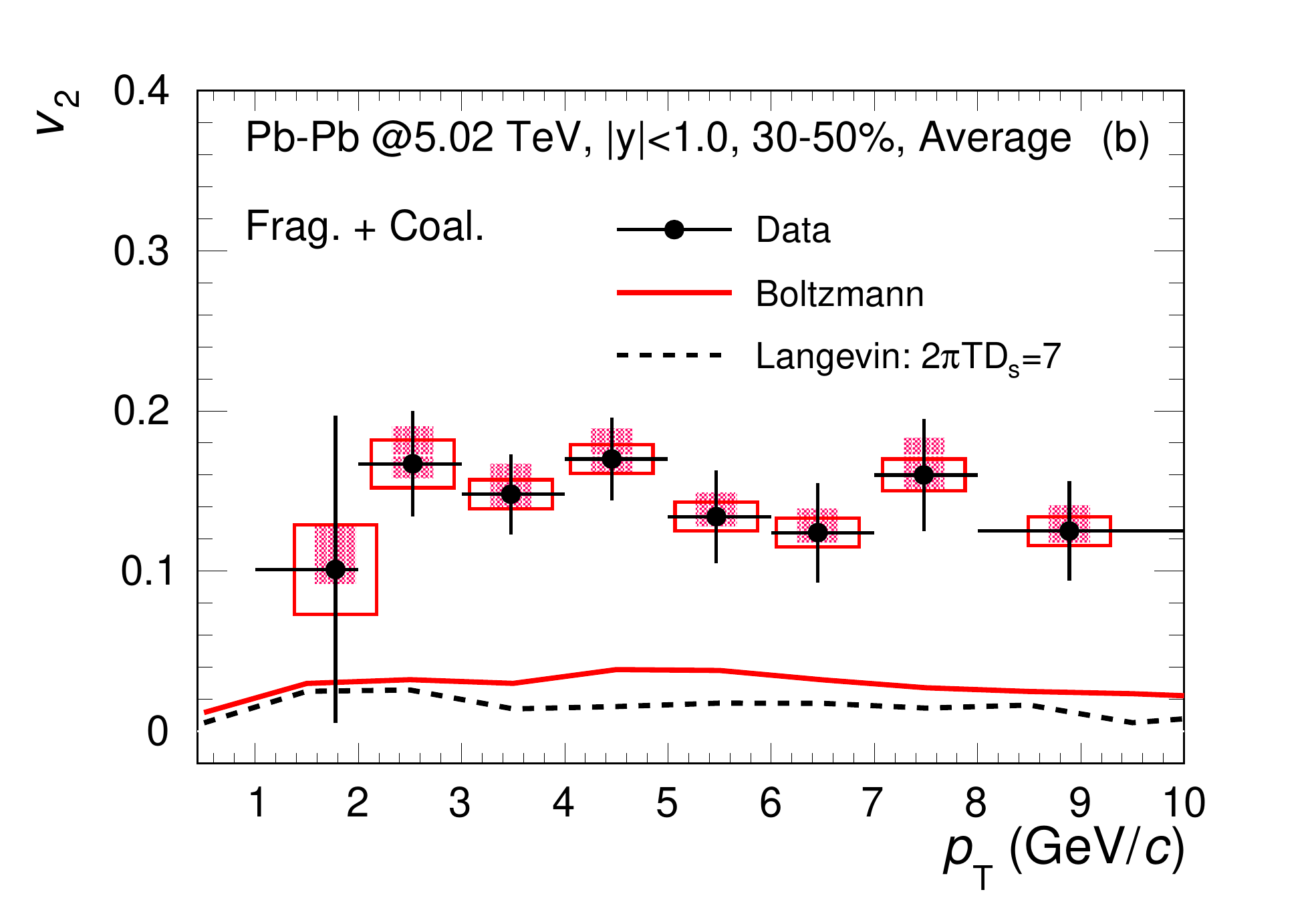}
\caption{(Color online) (a) Comparison of the nuclear modification factor $\raa$ of nonstrange D-meson ($D^{0}$, $D^{+}$ and $D^{*+}$)
with the Boltzmann (solid red curve) and Langevin approach (dashed black curve),
in central ($0-10\%$) Pb--Pb collisions at $\snn=5.02~{\rm TeV}$.
(b) Same as (a) but for $\vtwo$ obtained in semicentral ($30-50\%$) Pb--Pb collisions at $\snn=5.02~{\rm TeV}$.
Experimental data taken from Ref.~\cite{ALICEDesonPbPb5020RAA, ALICEDesonPbPb5020V2}.
Only the collisional energy loss mechanism is considered.}
\label{fig:DRAAV25020_Col}
\end{center}
\end{figure}

Figure~\ref{fig:DRAAV25020_Col} presents the average $\raa$ [(a)] and $\vtwo$ [(b)]
of the nonstrange D-meson ($D^{0}$, $D^{+}$ and $D^{*+}$) in central ($0-10\%$)
and semicentral ($30-50\%$) Pb--Pb collisions at $\snn=5.02~{\rm TeV}$, respectively,
with the Boltzmann (solid red curve) and Langevin approach (dashed black curve).
It is observed that $\raa$ is suppressed at high $\pt$ for the Boltzmann dynamics as compared to the Langevin,
while $\vtwo$ is systematically higher in the while $\pt$ region.
This behavior is consistent with the results found at parton level (see Fig.~\ref{fig:HQRAAV25020_Col}).
The available measurements for $\raa$ and $\vtwo$ (boxes) are shown for comparison.
The calculations with both the Boltzmann and Langevin approaches fail to
reproduce both the $\raa$ and $\vtwo$ data,
which may due to the missing effects such as radiative energy loss mechanisms.

Based on the Bayesian model-to-data analysis,
the original Lido hybrid model~\cite{Lido18} is developed to study
the fundamental interaction mechanisms between heavy quark and the QGP constituents.
However, in this work, we utilize only its Boltzmann module
to describe the charm quark propagation inside the underlying thermal medium.
Therefore, one cannot expect same $\raa$ and $\vtwo$ results between us,
since the other used modules are different such as the initial charm quark momentum spectra, hydrodynamic modeling
and the heavy-light coalescence in the subsequent hadronization procedure.
\section{Boltzmann and Langevin Dynamics with both elastic and inelastic processes}\label{sec:Results_ColRad}
Concerning the scattering inelastically with the light (anti-)quarks and gluons of QGP in Boltzmann,
both the $2\rightarrow 3$ gluon radiation and the $3\rightarrow 2$ inverse absorption processes
are taken into account to guarantee the detailed balance.
The relevant scattering matrixes are derived in an improved Gunion-Bertsch model in the Soft-Eikonal limit~\cite{BAMPS10}.
Meanwhile, a Debye screening mass $m_{\rm D}^{2}=\frac{8}{\pi}(N_{c}+N_{f})\alpha_{s}T^{2}$
based on the Boltzmann statistics~\cite{XuPRC17}
is considered to regulate the $t-$channel gluon prpagator.
The LPM effect is included by restricting the momentum space integration
of the emission-absorption gluon with a coherence factor~\cite{Lido18},
\begin{equation}
\begin{aligned}\label{eq:BTE_LPM}
\frac{d^{3}\vec{k}}{2k} \rightarrow \frac{d^{3}\vec{k}}{2k} \cdot \biggr\{ 2[1-cos\biggr(\frac{t-t_{0}}{\tau_{f}}\biggr)] \biggr\},
\end{aligned}
\end{equation}
where, $t_{\rm 0}$ is the initial time for gluon radiation/absorption,
$k_{\perp}$ is the transverse momentum of gluon,
$\tau_{f}$ the gluon formation time
\begin{equation}
\begin{aligned}\label{eq:BTE_Tauf}
\tau_{f}=\frac{2x(1-x)E}{k^2_{\perp}+(xm_{\rm Q})^2+(1-x)m_{\rm D}^{2}/2},
\end{aligned}
\end{equation}
with $E$ and $m_{\rm Q}$ are the HQ energy and mass, respectively, and
\begin{equation}
\begin{aligned}\label{eq:BTE_X}
x=\frac{k+k_{z}}{E+p_{z}}
\end{aligned}
\end{equation}
characterize the light-cone momentum fraction of the radiated/sbsorbed gluon.
Note that the coherence factor shown in Eq.~\ref{eq:BTE_LPM},
is obtained by requiring the radiation rate reduces to the Higher-Twist prediction
in the limits~\cite{Lido18}: soft-emission ($x\ll1$); large gluon transverse momentum comparing
with the momentum transfer ($k^2_{\perp}\gg q^2_{\perp}$).

The gluon radiation incorporated Langevin transport model is expressed as~\cite{CaoPRC15, CTGUHybrid1,CTGUHybrid2}
\begin{equation}
\begin{aligned}\label{eq:LTE_ColRad}
\frac{dp^{\rm i}}{dt}=F^{\rm i}_{\rm Drag} + F^{\rm i}_{\rm Diff} + F^{\rm i}_{\rm Gluon}.
\end{aligned}
\end{equation}
Comparing with Eq.~\ref{eq:LTE_Col},
the additional term $F^{\rm i}_{\rm Gluon}$ is the recoil force induced by the emitted gluons
\begin{equation}\label{eq:RecoilForce}
F^{\rm i}_{\rm Gluon}=-\frac{dp^{\rm ij}_{\rm Gluon}}{dt},
\end{equation}
where, $p^{\rm ij}$ indicates the momentum of the radiated gluon.
The transverse momentum together with the radiation time dependence of the radiated gluon
is quantified by pQCD Higher-Twist model~\cite{HTPRL04}:
\begin{equation}\label{eq:HigherTwist}
\frac{dN^{\rm Gluon}}{dz dk_{\perp}^{2} dt}=\frac{2\alpha_{s}C_{\rm A}P(x) \hat{q}_{\rm q}}{\pi k_{\perp}^{4}}
\biggr[\frac{k_{\perp}^{2}}{k_{\perp}^{2}+(xm_{\rm Q})^{2}}\biggr]^{4} sin^{2}\biggr( \frac{t-t_{\rm 0}}{2\tau_{\rm f}} \biggr).
\end{equation}
$x=k/E$ denotes the fraction of energy carried away by the emitted gluon,
which is equivalent to Eq.~\ref{eq:BTE_X} in the high-energy ($E\sim p_{z}$) and collinear-radiation ($k\sim k_{z}$) limt;
$\alpha_{s}(k_{\perp})=\frac{4\pi}{11N_{c}/3-2N_{f}/3}({\rm ln}\frac{k_{\perp}^{2}}{\Lambda^{2}})^{-1}$
is the strong coupling constant of QCD at leading order approximation;
$P(x)=(x^{2}-2x+2)/x$ is the splitting function for process ``$Q\rightarrow Q+g$'';
$\hat{q}_{\rm q}$ is the jet transport coefficient for
quarks\footnote[5]{According to its definition, $\hat{q}_{\rm q}=2\kappa_{\perp}/v_{\rm Q} \approx 2\kappa_{\perp}$
at high energy $E \gg m_{\rm Q}$, where HQ velocity $v_{Q}=\sqrt{1-(m_{\rm Q}/E)^{2}} \sim 1.$};
$\tau_{\rm f}=2x(1-x)E/[k_{\perp}^{2}+(xm_{\rm Q})^{2}]$ is the gluon formation time
without considering the contribution of the gluon thermal mass ($m^2_{g}=m_{\rm D}^{2}/2$; see Eq.~\ref{eq:BTE_Tauf}).
It was argued~\cite{CaoPRC15} that an additional lower cutoff was imposed on the emitted gluon energy, $k\geqslant\pi T$,
to balance the gluon radiation and the inverse absorption,
so as to ensure that HQ equilibrium state can be reached after sufficiently long evolution time.

We can see that the implementations of radiative energy loss are different in the Boltzmann and Langevin approaches,
which will apparently introduce the source of uncertainty when comparing these two models.
However, it is still necessary to check further the modifications for each dynamics.

\begin{figure}[!htbp]
\begin{center}
\vspace{-1.0em}
\setlength{\abovecaptionskip}{-0.1mm}
\setlength{\belowcaptionskip}{-1.5em}
\includegraphics[width=.46\textwidth]{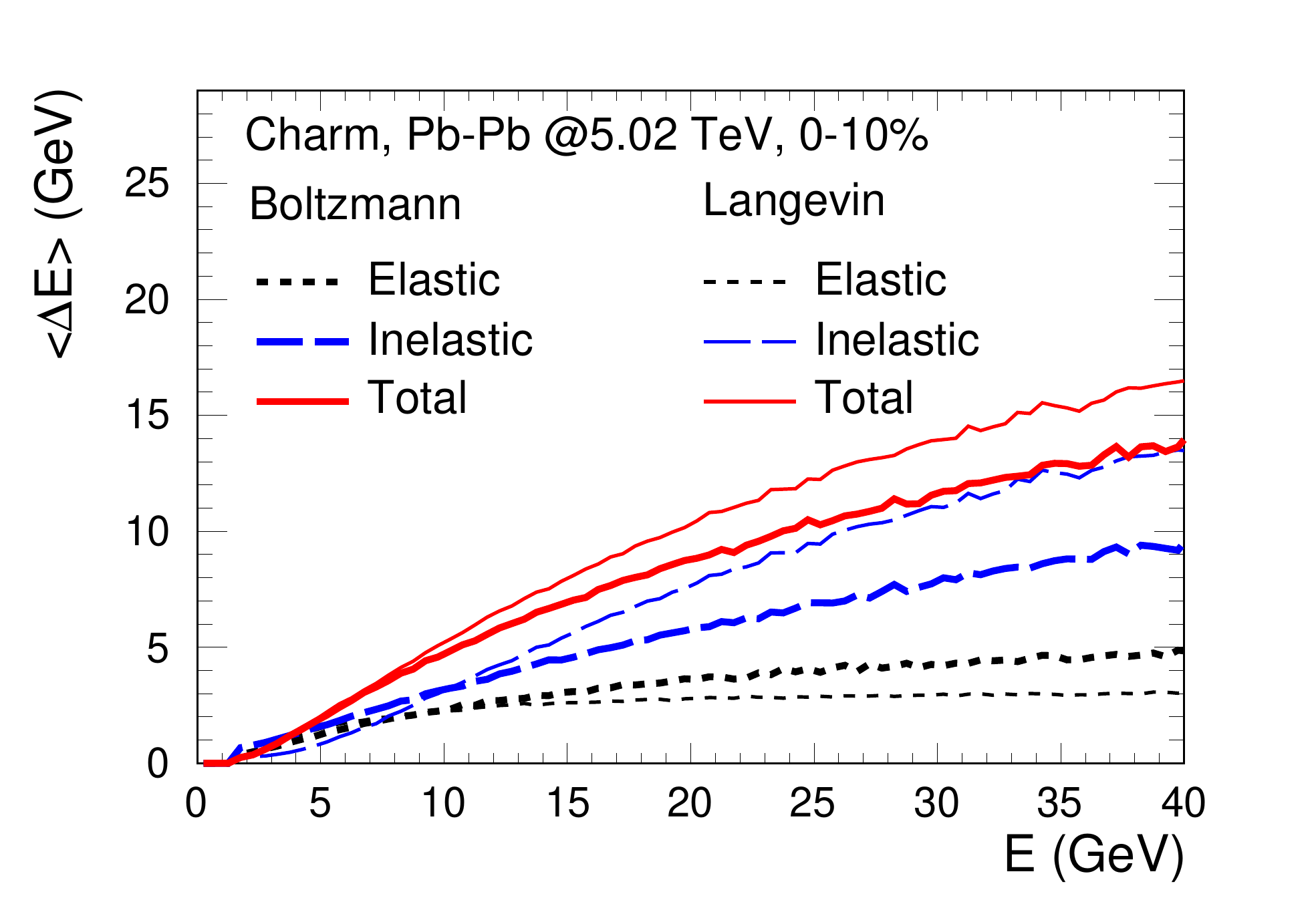}
\caption{Same as Fig.~\ref{fig:Eloss_Col} but including both the elastic and inelastic contributions.}
\label{fig:Eloss_ColRad}
\end{center}
\end{figure}

Figure~\ref{fig:Eloss_ColRad} displays the elastic (or collisional) and inelastic (or radiative) energy loss
as the dashed black and long dashed blue curves, respectively,
while the combined results are shown as the solid red curves.
The results with the Boltzmann and Langevin dynamics are presented
as the thick and thin curves, respectively.
We can see that the inelastic contribution (thick long-dashed blue curve) with the Boltzmann approach, is dominated at high energy,
while the elastic component (thick dashed black curve) is significant at low energy.
Similar behavior is observed with the Langevin approach (thin curves).
The energy loss due to elastic scattering is larger with Boltzmann (similar with Fig.~\ref{fig:Eloss_Col}),
while the total in-medium energy loss is larger with Langevin, in particular at high energy region.

\begin{figure}[!htbp]
\begin{center}
\vspace{-1.0em}
\setlength{\abovecaptionskip}{-0.1mm}
\setlength{\belowcaptionskip}{-1.5em}
\includegraphics[width=.46\textwidth]{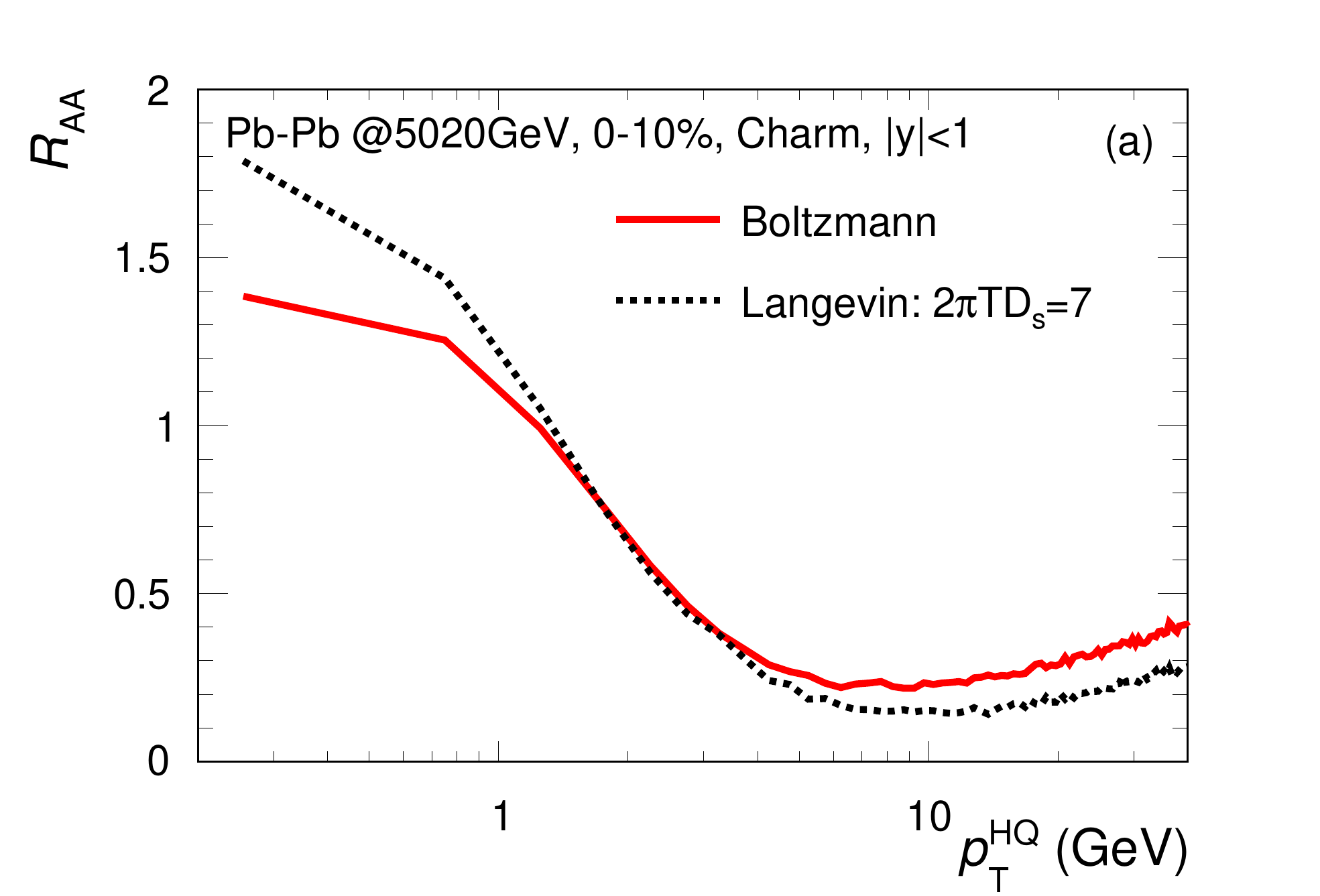}
\includegraphics[width=.46\textwidth]{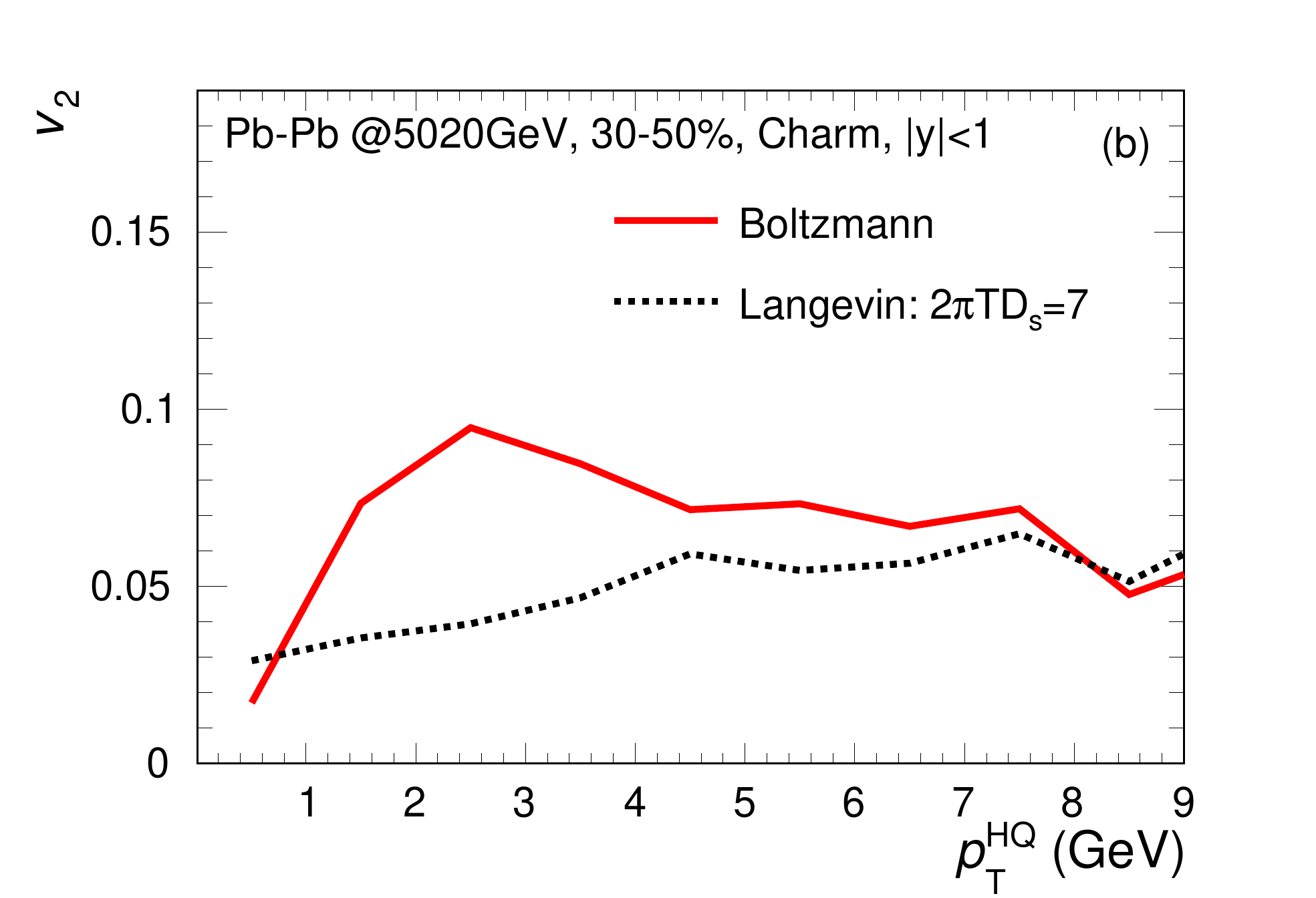}
\caption{Same as Fig.~\ref{fig:HQRAAV25020_Col} but including both the elastic and inelastic contributions.}
\label{fig:HQRAAV25020_ColRad}
\end{center}
\end{figure}

Figure~\ref{fig:HQRAAV25020_ColRad} shows the $\raa$ ($\vtwo$) of charm quark
with Boltzmann (solid red curves) and Langevin approach (dashed black curve) but including both
the elastic and inelastic scattering processes.
When comparing with the results including only the elastic component (see Fig.~\ref{fig:HQRAAV25020_Col}),
$\raa$ is suppressed (enhanced) at high (low) $\pt$ for both these two models, 
while $\vtwo$ is enhanced in the range $2\lesssim\pt\lesssim7~{\rm GeV}$, in particular with the Boltzmann approach.
This is mainly due to the fact that, as discussed in Fig.~\ref{fig:Eloss_ColRad},
inelastic component dominates at high $\pt$,
meanwhile, it introduces more interactions between charm quarks and QGP partons,
transfering more $\vtwo$ from QGP partons to charm quarks.
Similar behavior was observed in Ref.~\cite{DukeNPA13}.

\begin{figure*}[!htbp]
\begin{center}
\vspace{-1.0em}
\setlength{\abovecaptionskip}{-0.1mm}
\setlength{\belowcaptionskip}{-1.5em}
\includegraphics[width=.32\textwidth]{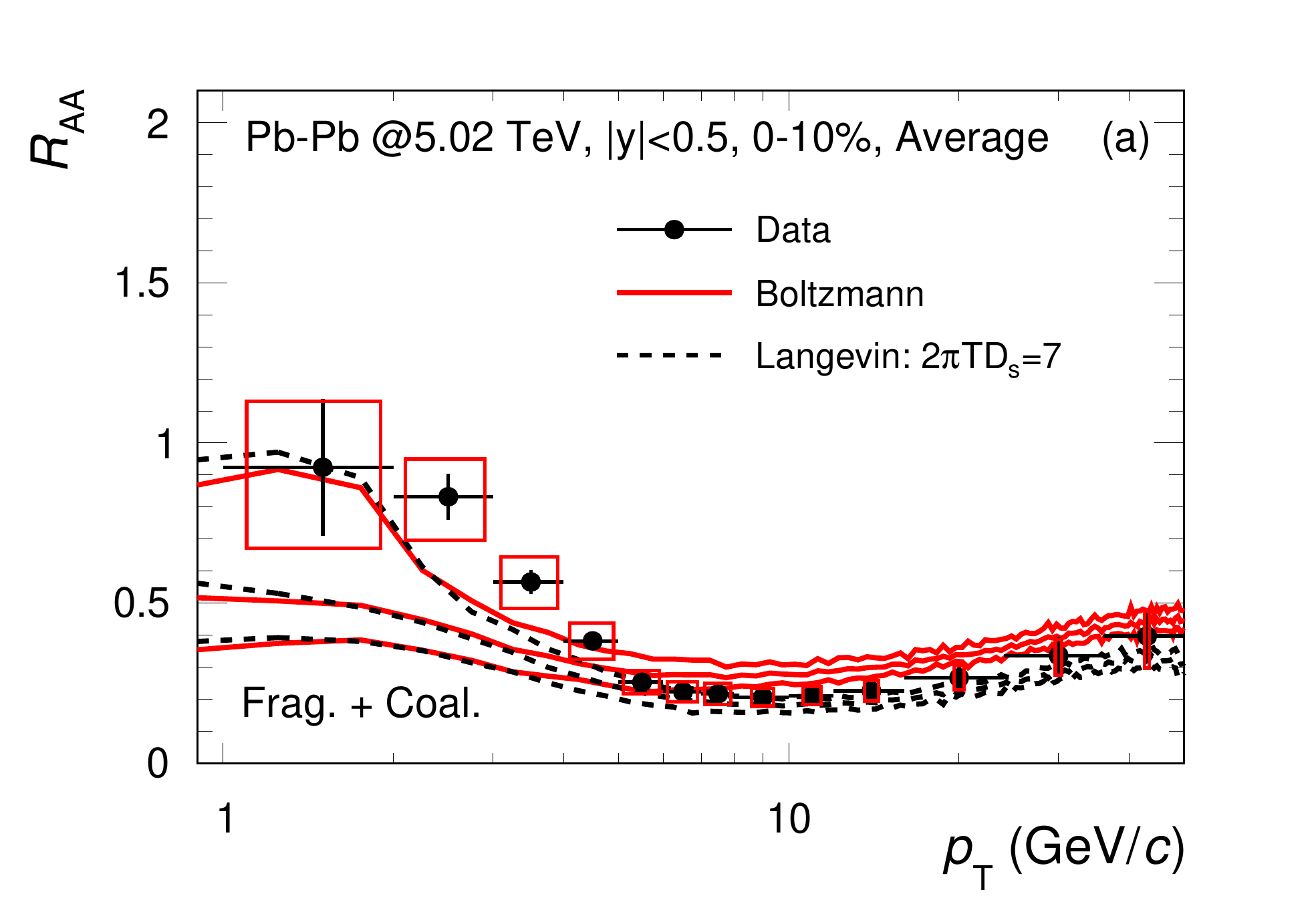}
\includegraphics[width=.32\textwidth]{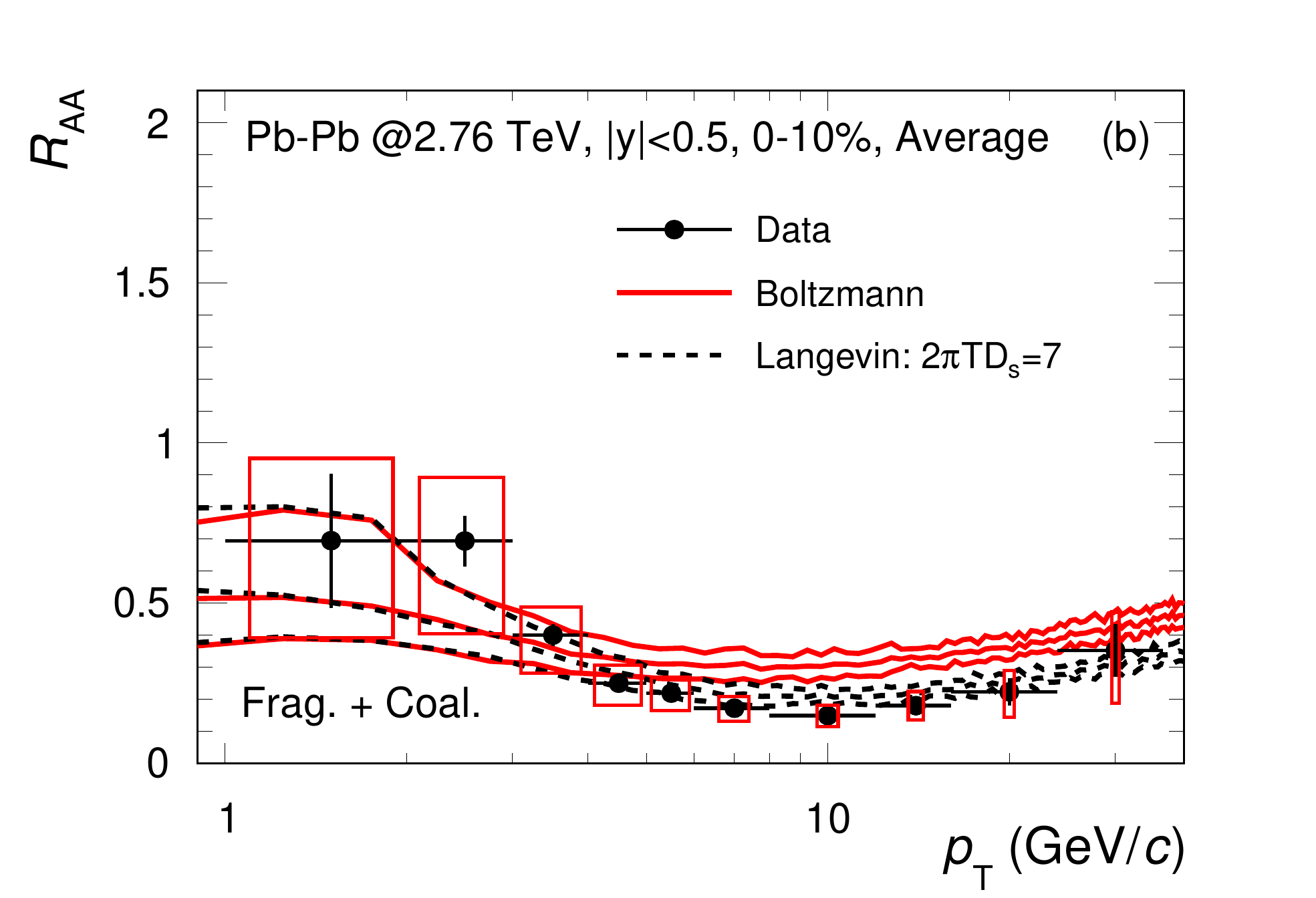}
\includegraphics[width=.32\textwidth]{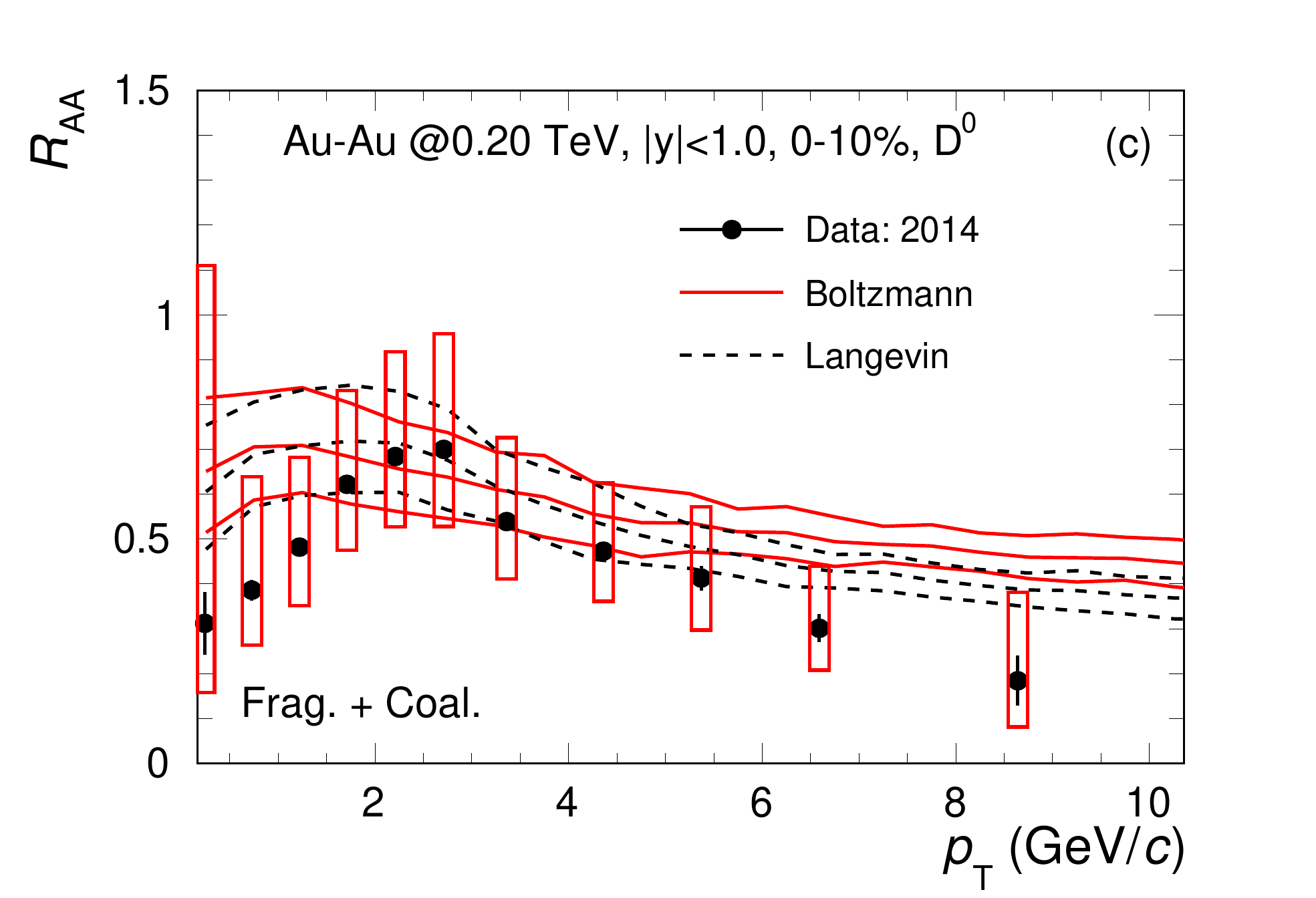}
\includegraphics[width=.32\textwidth]{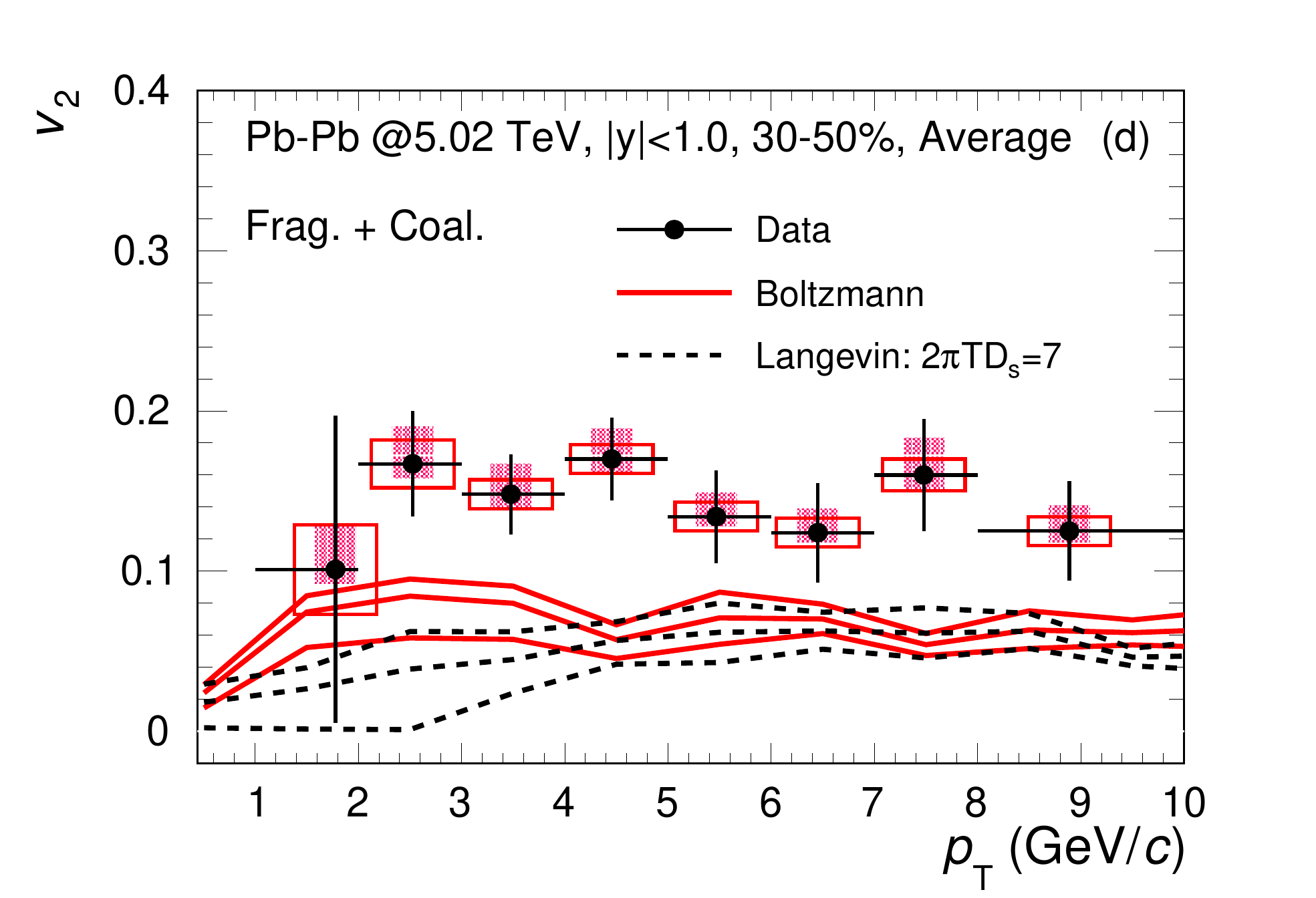}
\includegraphics[width=.32\textwidth]{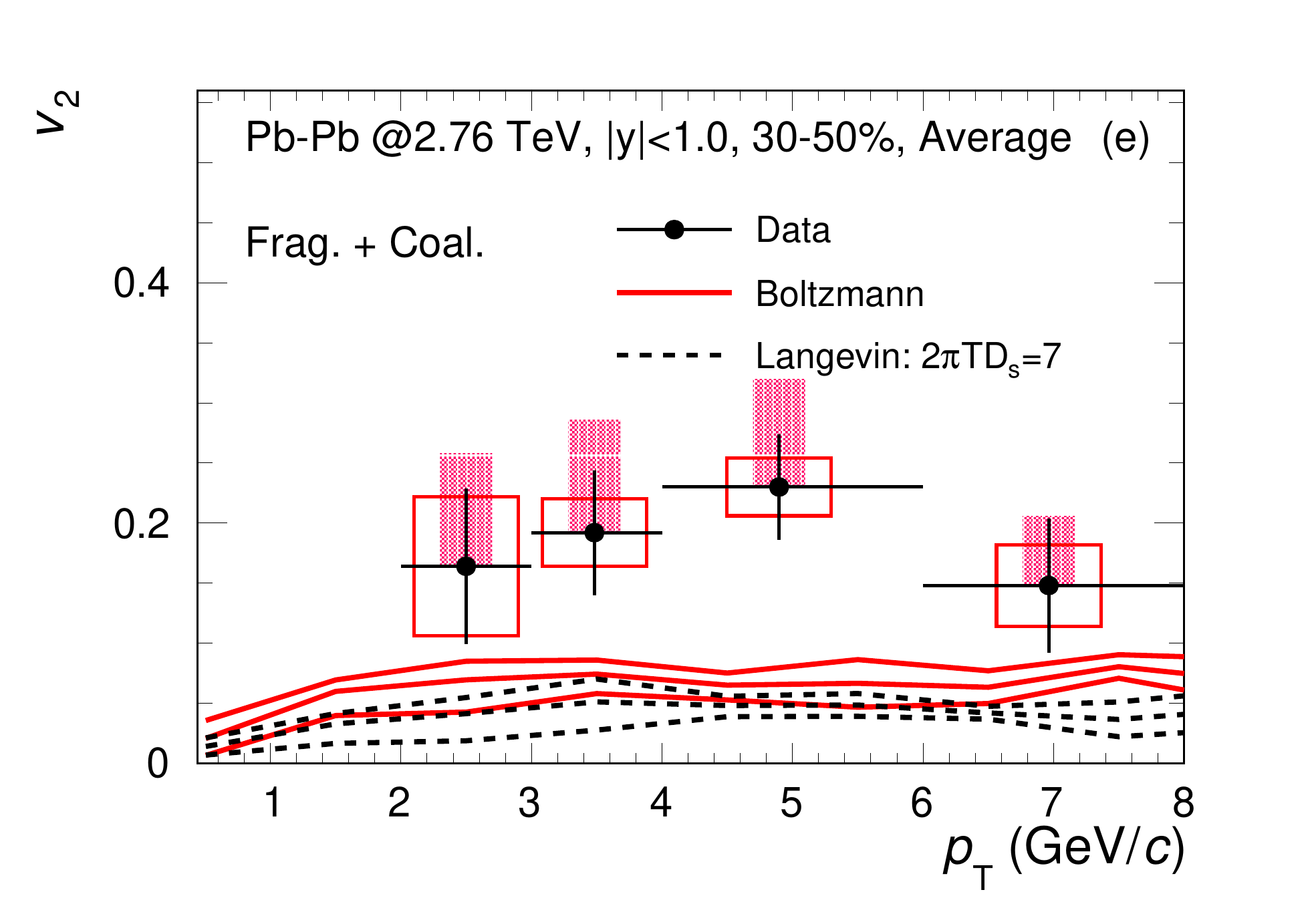}
\includegraphics[width=.32\textwidth]{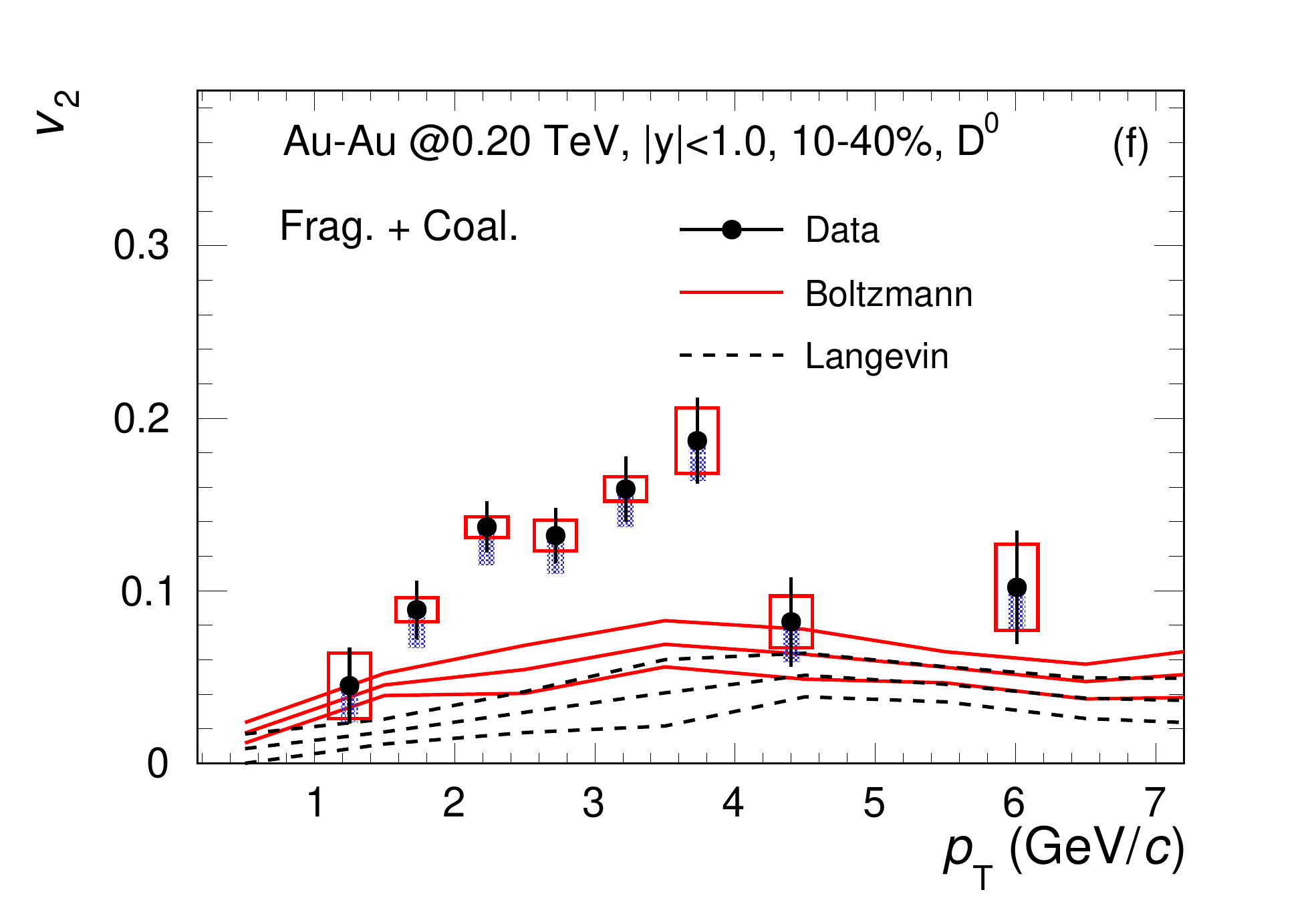}
\caption{Same as Fig.~\ref{fig:DRAAV25020_Col} but including both the elastic and inelastic contributions,
as well as the theoretical uncertainties:
[(a), (d)] nonstrange D-meson in Pb--Pb collisions at $\snn=5.02~{\rm TeV}$.
[(b), (e)] nonstrange D-meson in Pb--Pb collisions at $\snn=2.76~{\rm TeV}$;
[(c), (f)] $D^{0}$ in Au--Au collisions at $\snn=200~{\rm GeV}$.
Experimental data taken from Refs.~\cite{ALICEDesonPbPb5020RAA, ALICEDesonPbPb5020V2,
ALICEDesonPbPb2760RAA, ALICEDesonPbPb2760V2, STARD0AuAu200RAA, STARD0AuAu200V2}.}
\label{fig:DRAAV2Comb_ColRad}
\end{center}
\end{figure*}
Figure~\ref{fig:DRAAV2Comb_ColRad} presents the average $\raa$ [(a)] and $\vtwo$ [(d)]
of the nonstrange D-meson ($D^{0}$, $D^{+}$ and $D^{*+}$) in central ($0-10\%$)
and semicentral ($30-50\%$) Pb--Pb collisions at $\snn=5.02~{\rm TeV}$, respectively,
with the Boltzmann (solid red curves) and Langevin approach (dashed black curves).
The central values are obtained in terms of the central predictions of the initial heavy quark spectra and the nPDFs.
The bands are determined according to the total theoretical uncertainties,
which are contributed by the FONLL model predictions on the initial charm momentum spectra,
as well as the EPS09 NLO parameterization for the nPDF in Pb~\cite{CTGUHybrid1}.
We take the maximum derivation with respect to the central values in each $\pt$ bin,
and add in quadrature the above two components to get a conservative range.
It is observed that $\raa$ is suppressed at high $\pt$ for the Langevin dynamics as compared to the Boltzmann,
while $\vtwo$ is systematically higher at moderate $\pt$ ($2\lesssim\pt\lesssim4~{\rm GeV}$).
This behavior is consistent with the results found at parton level (see Fig.~\ref{fig:HQRAAV25020_ColRad}).
The available measurements for $\raa$ and $\vtwo$ (boxes) are shown for comparison.
We find that, (1) when comparing with Fig.~\ref{fig:DRAAV25020_Col},
both the elastic (or collisional) and inelastic (or radiative) energy loss mechanisms are needed to
reduce the discrepancy between model and data, as concluded in Ref.~\cite{GyulassyNPA07};
(2) the calculations with the Langevin approach seem to give a better description of the measured $\raa$
as compared to those with Boltzmann approach, in particular in the range $\pt\gtrsim10~{\rm GeV}$.
Meanwhile, nonstrange D-meson $\vtwo$ calculated with the Boltzmann approach
is closer to the available data at $\pt\lesssim4~{\rm GeV}$.
The comparison of $\raa$ and $\vtwo$ gives the opposite indications about the two models,
confirming that it is challenging to describe well $\raa$ and $\vtwo$ simultaneously, as observed in Ref.~\cite{Das15}.
A similar behavior can be observed in Pb--Pb collisions at $\snn=2.76~{\rm TeV}$ (panel-b and e in Fig.~\ref{fig:DRAAV2Comb_ColRad})
and Au--Au collisions at $\snn=200~{\rm GeV}$ (panel-c and f in Fig.~\ref{fig:DRAAV2Comb_ColRad}).
\section{Conclusion and Discussion}\label{sec:Summary}
In this work, we investigated the charm quark evolution via the Boltzmann and Langevin dynamics
in relativistic heavy-ion collisions.
By including only the elastic scattering contributions,
the extracted drag coefficient ($\eta_{\rm D}$), momentum diffusion coefficients ($\kappa_{\rm L}$ and $\kappa_{\rm T}$)
and spatial diffusion coefficient ($2\pi TD_{s}$) are calculated
as a function of charm quark energy and the medium temperature,
and further compared between the two approaches.
The relevant in-medium energy loss together with its effect on
the nuclear modification factor ($\raa$) and elliptic flow coefficient ($\vtwo$) at parton and hadron level,
are discussed and compared with the available measurements at RHIC and LHC energies.

It is found that $\eta_{\rm D}$, $\kappa_{\rm L}$ and $\kappa_{\rm T}$ calculated from the
Boltzmann dynamics ($2\pi TD_{s}\lesssim7$ in the range $1<T/T_{c}<3$),
are systematically larger than the ones obtained with the Langevin approach ($2\pi TD_{s}=7$).
The collisional energy loss is larger with the Boltzmann approach,
resulting in a smaller charm quark $\raa$ at $\pt\gtrsim10~{\rm GeV}$, as compared to the Langevin.
Meanwhile, due to the larger drag force and stronger interactions in Boltzmann,
it is more efficient in producing larger $\vtwo$,
as well as in developing the broadening effect for the azimuthal angle distributions.
The above $\raa$ and $\vtwo$ behaviors observed at parton level are well inherited by the corresponding heavy-flavor hadrons.
When comparing the model with available data,
it is realized that the calculations including only the contributions from the elastic processes,
are unable to describe both the $\raa$ and $\vtwo$ measured at RHIC and LHC energies.
This discrepancy can be reduced by including the inelastic contributions in both the Boltzmann and Langevin dynamics,
even though the relevant implementations are different between these two models.
Finally, we find that the model calculations for non-strange D-meson $\raa$ favor the Langevin approach,
while $\vtwo$ prefer the Boltzmann approach.
A simultaneous description of both  $\raa$ and $\vtwo$ remains a challenge for both models. 

It is necessary to mention that Ref.~\cite{HQBoltLang14} is also a systematical study of Boltzmann versus Langevin,
by considering only the elastic scattering processes.
We obtain the similar conclusions for charm quarks, for instance,
(1) drag coefficients show a decreasing momentum/energy dependence,
while the momentum diffusion coefficients present an increasing behavior  from the Boltzmann transport equation;
(2) after the in-medium evolution,
charm quark spectra is harder with the Boltzmann approach in the range $\pt\lesssim7-10~{\rm GeV}$,
resulting in a larger (smaller) $\raa$ at $2\lesssim\pt\lesssim7~{\rm GeV}$ ($\pt\lesssim2~{\rm GeV}$);
(3) as explained above, Boltzmann model gives larger $\vtwo$ at both parton and hadron levels.
On the other hand, few differences are observed between us:
(1) the calculations for $\raa$ and $\vtwo$ with both the Boltzmann and Langevin dynamics,
including only the elastic processes,
are failed to describe the available data in this analysis,
while it is not true in Ref.~\cite{HQBoltLang14}, in particular for the Boltzmann approach,
which reproduce well the measured $\pt$ dependence of both $\raa$ and $\vtwo$ (see references therein);
(2) in this analysis, the additional inelastic (or radiative) contributions
are powerful to reduce the discrepancy with data, in particular at $\pt$,
however, this effect is not discussed in Ref.~\cite{HQBoltLang14}; 
(3) the theoretical uncertainty such as the one on the initial charm quark production,
is taken into account in this work, which is missing in Ref.~\cite{HQBoltLang14}.
These differences could be induced by the following sources:
(1) comparing with the hybrid model utilized in this analysis (Sec.~\ref{sec:Method}),
Ref.~\cite{HQBoltLang14} takes different approaches in the relevant modules,
such as the initial charm quark spectra is given by a parameterized power-law function,
which works better only at high momentum region;
nuclear (anti-)shadowing and heavy-light coalescence effects are missing;
(2) with the Boltzmann approach,
only quark-gluon scattering ($Q+g\rightarrow Q+g$) is considered in Ref.~\cite{HQBoltLang14},
while quark-quark ($Q+q\rightarrow Q+q$) is also included in the two-body interactions in this work;
constant running coupling and Debye mass are used in Ref~\cite{HQBoltLang14},
but a momentum and temperature dependent scenario is adopted for us.

Finally, it is interesting to note that
the resolution of the above model-to-data challenge may require the inclusion of nonperturbative dynamics in the medium.
It may be noted that a similar challenge was previously investigated for
light flavor jet energy loss and a viable solution was previously proposed by
introducing a nontrivial medium color structure that includes both chromo-electric
and chromo-magnetic degrees of freedom~\cite{LiaoPRL08, LiaoPRL09}
and that leads to a strong temperature dependence of transport coefficients~\cite{CUJET3CPC18, CUJET3JHEP16,CUJET3Arxiv18}.
Whether a similar strategy may help address the $\raa$ and $\vtwo$ challenge in the heavy flavor sector
would be an interesting problem for future investigation.
More detailed studies will be reported in forthcoming publications.
\begin{acknowledgments}
The authors are grateful to Weiyao Ke for providing the Boltzmann module in Duke model,
as well as the data shown in Fig.~\ref{fig:DsVsT}.
S.~Li is supported by National Science Foundation of China (NSFC) under Grant Nos.11847014 and 11875178,
China Three Gorges University (CTGU) Contracts No.1910103,
Hubei Province Contracts No.B2018023, 
China Scholarship Council (CSC) Contract No.201807620007,
and the Key Laboratory of Quark and Lepton Physics Contracts No.QLPL2018P01.
C.~W.~Wang acknowledges the support from the NSFHB No.2012FFA085.
R.~Z.~Wan acknowledges support from NSFC under Project No.11505130.
J.~F.~Liao is supported by the National Science Foundation under Grant No.PHY-1352368.
The computation of this research was performed on IU's Big Red II cluster,
which was supported in part by Lilly Endowment, Inc., through its support for the
Indiana University Pervasive Technology Institute, and in part by the Indiana METACyt Initiative.
The Indiana METACyt Initiative at IU was also supported in part by Lilly Endowment, Inc.
\end{acknowledgments}

%
%
%

%
\end{document}